\documentclass[draft]{agujournal2019}
\usepackage{url} 
\usepackage{lineno}
\usepackage{soul}
\draftfalse

\journalname{arXiv}

\usepackage{tikz,pgfplots}
\usepackage{amsmath}
\pgfplotsset{compat=1.7}

\DeclareRobustCommand{\ColorCircle}[1]{{\tikz{\node[mark size=3pt,color=#1] at (0,0) {\pgfuseplotmark{*}};}}}
\DeclareRobustCommand{\ColorSquare}[1]{{\tikz{\node[mark size=3pt,color=#1] at (0,0) {\pgfuseplotmark{square*}};}}}

\newcommand\solidrule[1][0.25cm]{\rule[0.5ex]{#1}{2pt}}
\newcommand\dashedrule{\mbox{%
 \solidrule[2mm]\hspace{2mm}\solidrule[2mm]}}

\newcommand{\dotrule}{%
\mbox{%
 \solidrule[0.5mm]\hspace{0.75mm}\solidrule[0.5mm]\hspace{0.75mm}\solidrule[0.5mm]\hspace{0.75mm}\solidrule[0.5mm]}} 
	
\DeclareRobustCommand{\BoundSquare}[4]{\tikz{\draw [#1,line width=#2,draw=#3,fill=#4] (0,0) rectangle (0.75em,0.75em);}}	

\DeclareRobustCommand{\BoundCircle}[4]{\tikz{\draw [#1,line width=#2,draw=#3,fill=#4] (0,0) circle (0.375em);}}

\DeclareRobustCommand{\FlowInput}{{\tikz{\draw[] (0,0) circle (0.35em)}}}
\DeclareRobustCommand{\FlowModelComp}{{\tikz{\draw[] (0,0) rectangle (0.9em,0.6em)}}}
\DeclareRobustCommand{\FlowModel}{{\tikz{\draw[line width=1.5pt] (0,0) rectangle (0.9em,0.6em)}}}
\DeclareRobustCommand{\FlowOutput}{{\tikz{\draw[rounded corners=0.23em] (0,0) rectangle (0.9em,0.6em)}}}

\definecolor{OWQ_1}{rgb}{0.3467 ,0.5360,0.6907}%
\definecolor{OWQ_2}{rgb}{0.9153 ,0.2816 , 0.2878}%
		
\definecolor{PRSA_1}{rgb}{0.34667,0.53600,0.69067}%
\definecolor{PRSA_2}{rgb}{0.91529,0.28157,0.28784}%
\definecolor{PRSA_3}{rgb}{0.44157,0.74902,0.43216}%
\definecolor{PRSA_4}{rgb}{1.00000,0.59843,0.20000}%

\definecolor{ColMat_1}{rgb}{0.904706,0.191765,0.19882}
\definecolor{ColMat_2}{rgb}{0.294118,0.544706,0.749412}
\definecolor{ColMat_3}{rgb}{0.371765,0.717647,0.361176}
\definecolor{ColMat_4}{rgb}{1,0.548235,0.1}
\definecolor{ColMat_5}{rgb}{0.865,0.811,0.433}
\definecolor{ColMat_6}{rgb}{0.685882,0.403529,0.241176}
\definecolor{ColMat_7}{rgb}{0.971765,0.555294,0.774118}
\definecolor{ColMat_8}{rgb}{0.64,0.64,0.64}

\definecolor{grey_1}{rgb}{0.7,0.7,0.7}

\renewcommand{\Re}{\operatorname{Re}}

\begin{document}

%
%


\title{Model predictions of wave overwash extent into the marginal ice zone}

%
%



\authors{Jordan~P.~A.~Pitt\affil{1},
Luke~G.~Bennetts\affil{1},
Michael~H.~Meylan\affil{2},
Robert~A.~Massom\affil{3,4,5},
Alessandro~Toffoli\affil{6},}

\affiliation{1}{School of Mathematical Sciences, University of Adelaide, Adelaide, South Australia 5005, Australia}
\affiliation{2}{School of Information and Physical Sciences, The University of Newcastle, Callaghan, New South Wales 2308, Australia}
\affiliation{3}{Australian Antarctic Division, Kingston, Tasmania 7050, Australia}
\affiliation{4}{Australian Antarctic Program Partnership, Institute for Marine and Antarctic Studies, University of Tasmania, Hobart, Tasmania 7001, Australia.}
\affiliation{5}{The Australian Centre for Excellence in Antarctic Science, University of Tasmania, Hobart, Tasmania 7001, Australia}
\affiliation{6}{Department of Infrastructure Engineering, University of Melbourne, Parkville, Victoria 3052, Australia}

\correspondingauthor{Jordan~P.A.~Pitt}{jordan.pitt@adelaide.edu.au}

\begin{keypoints}
\item Propose model for the extent of wave overwash of ice floes in the marginal ice zone incorporating wave attenuation and floe--wave motions.
\item Wave overwash extents range from a few hundred metres for calm seas to over 3\,km for high seas.
\item For mean sea states, pancake floe fields typically experience wave overwash extents $75\%$ greater than fragmented floe fields.
\end{keypoints}

\begin{abstract}
A model of the extent of wave driven overwash into fields of sea ice floes is proposed. The extent model builds on previous work modelling wave overwash of a single floe by regular waves by including irregular incoming waves and random floe fields. The model is validated against a laboratory experiment.
It is then used to study the extent of wave overwash into marginal ice zones consisting of pancake and fragmented floe fields. The effects of wave conditions and floe geometry on predicted extents are investigated. Finally, the model is used to predict the wave overwash extent for the conditions observed during a winter (July) $2017$ Antarctic voyage in which the sea surface was monitored by a stereo-camera system.
\end{abstract}
\section*{Plain Language Summary}
Wave overwash is the flow of water across the surface of a floating body due to passing waves. It is an important process in the marginal ice zone where floating bodies of sea ice called floes and ocean waves interact. 
Overwash reduces the energy of waves, limiting their ability to travel deeper into the ice-covered ocean. Moreover, overwash removes snow and wets the surface of floes, which can drive growth or melting. 
Water left on the surface of a floe can host organisms and affects chemical processes taking place there. Overwash is pervasive in laboratory experiments where it has been measured. 
In contrast, there are scarce reports of overwash in the field and the extent of its occurrence has not been measured.
This paper proposes a model predicting the region of the marginal ice zone in which overwash occurs. The model is then validated against a laboratory experiment. 
The model is used to predict overwash extent for fields of pancake and fragmented floes, for calm to high wave conditions, including the conditions observed during a winter voyage in the Antarctic marginal ice zone, where overwash could have been observed. 

\section{Introduction}
\label{sec:intro}
The marginal ice zone (MIZ) is the highly dynamic outer band of the sea ice covered ocean in both polar regions, in which surface gravity waves and sea ice interact. The sea ice cover in the MIZ varies between regions \cite{Weeks-2010} and depends on the season \cite{Doble-2003,Toyota-2011,Hwang-2017,Alberello-2019} and synoptic conditions \cite{Vichi-2019,Finocchio-2020}. The ice cover is comprised of floating bodies of consolidated sea ice called floes \cite{Armstrong-1956} separated by interstitial ice or open water. The concentration (fraction of ice covered ocean surface) of floes and interstitial ice is typically combined into a single ice concentration \cite{Beitsch-2014}. Ice concentration typically increases with distance from the open (ice-free) ocean, which can occur gradually in a diffuse ice edge or suddenly in a compact ice edge \cite{Weeks-2010}. 
Bands of sea ice separated by open water can also form \cite{Wadhams-1983}. The ice cover in the MIZ can be broadly characterised as (i)~pancake floes \cite{Weeks-1982,Rothrock-1984,MIZEX-86,Doble-2003,Alberello-2019} or (ii)~fragmented floes and brash ice \cite{Toyota-2011,Frankenstein-2001,Collins-2015}. Type~(i) MIZs are formed when the surface of the ocean freezes with waves preventing full consolidation of the sea ice cover \cite{Wadhams-1987} whilst driving collisions of adjacent floes resulting in their distinctive `pancake' shape \cite{Weeks-1982,Shen-2001}. Type~(i) MIZs are thus generally associated with the autumn and winter advance of sea ice. Type~(ii) MIZs are produced by the break up of large floes due to wave-induced flexural stresses \cite{Prinsenberg-2011,Kohout-2016}. Type~(ii) MIZs can occur year round but are particularly associated with the retreat of the ice cover in spring and summer. 

Waves set each individual floe in the MIZ into motion, where the motion experienced by a particular floe depends both on its properties and those of the local wave field.
A floe sufficiently small relative to the local wavelength moves approximately in-phase with the waves and does not disturb the surrounding wave field other than to dissipate a small proportion of wave energy \cite{Squire-1995,Ardhuin-2016}. %
As the floe size relative to the local wavelength increases, the floe moves increasingly out of phase with the waves and begins to scatter wave energy in different directions \cite{Meylan-1994,Bennetts-2015}.
Waves attenuate with distance from the open ocean in the MIZ due to an accumulation of scattering and dissipation \cite{Squire-2020}, thus limiting the MIZ width and forming a coupled wave--floe interaction system. Short period components of the wave spectrum attenuate quickly and penetrate only several kilometres into the MIZ, whereas long-period components attenuate slowly and can penetrate tens to hundreds of kilometres into the MIZ \cite{Squire-1980,Meylan-2014}.
Therefore, the wave spectrum skews towards longer periods over distance \cite{Alberello-2021}.

\begin{figure}
\centering
\includegraphics[width=0.5\textwidth]{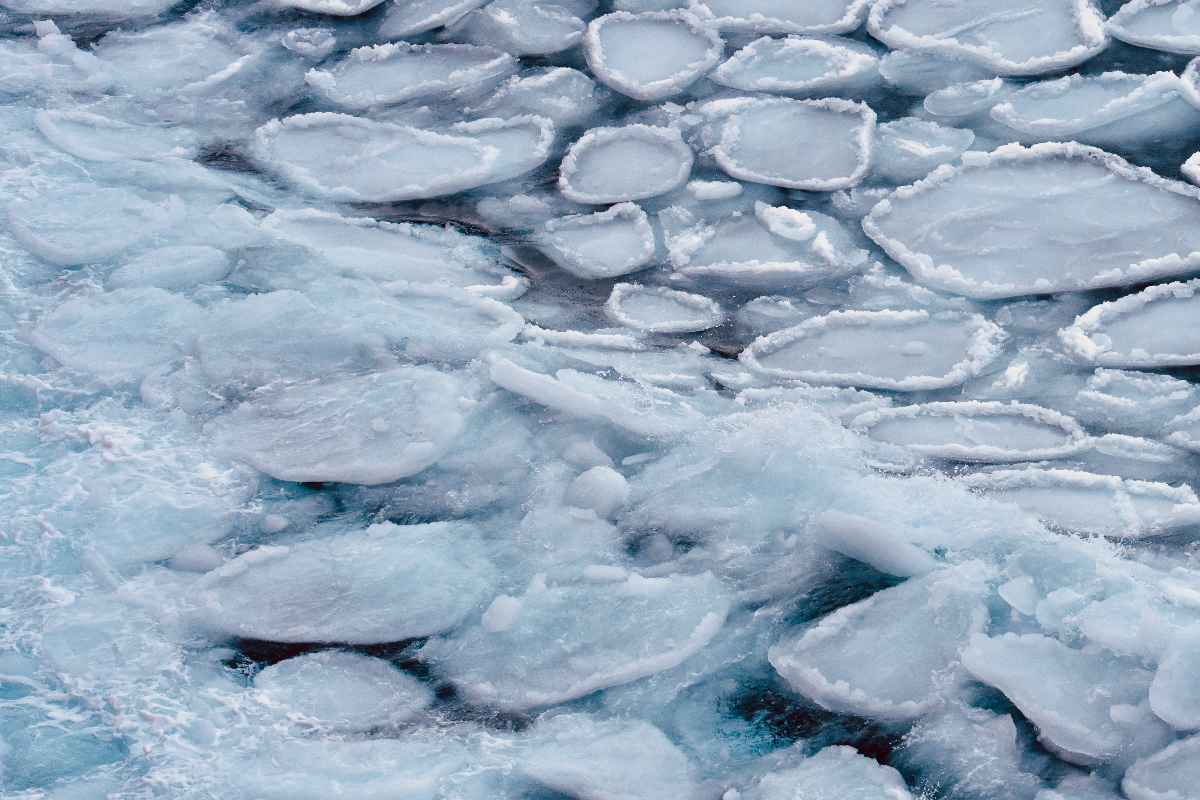}
\caption{Photo of pancake floes in the wake of the S.A.~Agulhas II on its winter 2019 voyage. The wake is propagating to the right and water deposited on the floe surfaces by overwash can be seen behind the leading wave, with some floes completely submerged.}
\label{Fig:OWExample}
\end{figure}

In the scattering regime, where floe and wave motions are out of phase, water can flow across the surface of a floe in a process known as wave overwash \cite{Massom-1997,Massom-1998} hereafter referred to as overwash. Overwash is a dynamic wave driven process distinct from flooding, which occurs where floes become submerged due to mass added by snow or other floes \cite{Wadhams-1987}.
Figure~\ref{Fig:OWExample} shows the overwashing of pancake floes by the wake of the S.A.~Agulhas~II during its winter 2019 voyage. The overwash produced by the wake deposited a significant amount of water on the encountered floes, submerging some floes completely. The process of overwash is the same whether waves are produced by the ship as in Figure~\ref{Fig:OWExample} and \citeA{Dumas-2021PP} or are naturally occurring as in the MIZ generally. While there are reports of overwash due to ocean waves in the MIZ \cite{Massom-1997,Massom-1998}, there has been no systematic study of overwash in the field. 
This can be largely explained by overwash almost certainly being most prevalent at the outermost region of the MIZ during large wave events, which would create very challenging test conditions.

Over the past ten years, laboratory experiments have become a common approach to modelling wave--floe interactions and wave attenuation in the MIZ, using either fresh water ice \cite{Dolatshah-2018,Yiew-2019,Alberello-2021a}, model
ice \cite{Cheng-2019,Passerotti-2022}, or, most often, artificial floes \cite{Montiel-2013a,Bennetts-2015,Bennetts-2015a, Meylan-2015,Toffoli-2015,Bai-2017,Nelli-2017,Yiew-2017,Sree-2018,Sree-2020,Huang-2022,Toffoli-22}. In contrast to the scarce reporting in the field, overwash is a pervasive feature of laboratory experiments.
In laboratory experiments overwash occurs even for relatively small steepness waves, which would usually be considered linear, due to the small freeboard of floes \cite{Skene-2015}.
Analysis of experimental data has given strong evidence that overwash causes wave energy dissipation and, hence, increases attenuation \cite{Bennetts-2015,Bennetts-2015a,Toffoli-2015,Nelli-2017,Toffoli-22}. Overwash has also been associated with accelerated melt of model ice and crumbling of the ice edge \cite{Passerotti-2022}.
The laboratory experiments have motivated development of theoretical \cite{Skene-2015,Skene-2018,Skene-2021} and numerical \cite{Huang-2019,Huang-2019a,Nelli-2020,TranDuc-2020} overwash models, which show promising agreement with measurements in terms of overwash properties and wave attenuation due to overwash, but are restricted to regular incoming waves and single floes at present.

In this article, recent progress in modelling overwash of a single floe by regular incoming waves \cite{Skene-2015} is built upon, to develop a stochastic model of overwash extent in the MIZ caused by irregular incoming waves. The model incorporates the ice conditions, described by an ice thickness, a floe concentration and a floe size distribution (FSD). 
The model includes wave attenuation due to scattering by the floes and an empirical model for wave dissipation in the MIZ. 
Thus, the model is capable of predicting the occurrence and extent of overwash into the MIZ for given wave and ice conditions. The model is validated using a laboratory experiment, for which overwash was previously reported but not quantified \cite{Bennetts-2015}. 
The model is then used to investigate the effect of calm and high seas in the Southern Ocean \cite{Young-2020} for the Antarctic MIZ when it is comprised of pancake floes \cite{Alberello-2019} or fragmented floes \cite{Toyota-2011}. Finally, the model is used to predict the region of overwash for inputs that replicate the wave and ice conditions during a field experiment in the Antarctic MIZ, where stereo-camera images of waves and ice floes were captured \cite{Vichi-2019,Alberello-2019,Alberello-2021}. 

\section{Model Overview}
\begin{figure}
\centering
\includegraphics[width=\textwidth]{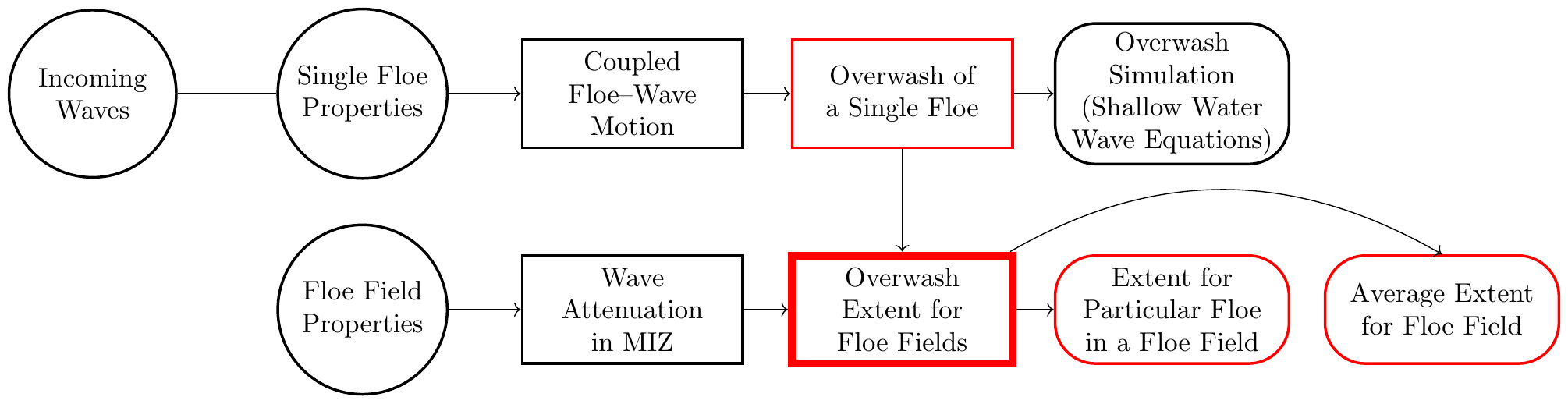}
\caption{Overview of model, showing inputs (\FlowInput{}), sub-models (\FlowModelComp{}), main model (\FlowModel{}) and outputs (\FlowOutput{}), with novel contributions highlighted in red.}
\label{Fig:Model_Flow}
\end{figure}

Figure~\ref{Fig:Model_Flow} gives an overview of the overwash extent model, divided into the inputs (circles; first and second columns), sub-models (rectangles; third and fourth columns),
main model (rectangle with thick borders; second row, fourth column), outputs (rectangles with rounded corners; fifth and sixth columns) and the relationships (arrows). 
The novel model contributions are highlighted in red.

In the first row, the overwash of a single floe with known properties by incoming waves is predicted. Ocean waves are irregular \cite{Holthuijsen-2010} and, thus, the model is stochastic, describing an ensemble of possible incoming waves for a given wave spectrum (see \S\ref{sec:Incoming_Waves}). 
The coupled floe--wave motion sub-model is linear and deterministic, providing the motions of the floe and waves in response to specified incoming waves. From the coupled floe--wave motions, the average frequency of overwash events of the single floe for the ensemble of incoming waves \cite{Rice-1945,Holthuijsen-2010} is calculated. The ensemble average frequency of overwash events of the single floe is then used to determine if the floe is overwashing, producing the nonlinear (amplitude dependent), stochastic sub-model for overwash of a single floe. 

In the second row, the overwash of a single floe is extended to a field of floes, assuming no floe--floe interactions (collisions or rafting). 
The floe fields add an extra layer to the stochastic nature of the model, as they are described by an ensemble of possible floe fields for a given floe size distribution (see \S\ref{subsubsec:FSD})
\cite{Rothrock-1984}. 
The ensemble average attenuation of irregular waves due to the ensemble of floe fields encountered is calculated by the linear wave attenuation sub-model \cite{Bennetts-2007,Meylan-2014}.
The attenuated irregular waves in the floe field are used to determine the overwash event frequency of individual floes in the floe field, thereby generating a nonlinear, stochastic model of overwash extent in the MIZ.
Two outputs are used to measure extent of overwash, i.e.\ the distance from the boundary of the floe field until overwash events for floes are sufficiently rare. The first gives the maximum distance a particular floe of known properties can be placed into a random floe field and be overwashed. The second gives the maximum distance most floes in the random floe field will be overwashed using the expected value of overwash event frequency for the random floe field.


\section{Model Components}\label{sec:Model_Comp}
\subsection{Overwash of a Single Floe}
\subsubsection{Incoming Waves}
\label{sec:Incoming_Waves}

The irregular incoming waves are described using a large number $N$ of regular wave components, which are equally spaced in some fixed frequency range $[\omega_1,\omega_N]$. Each regular wave component has a prescribed angular frequency $\omega_n$, a phase $\theta_n$ selected randomly from uniformly distribution over $[0,2\pi)$, and an amplitude $A_n$ selected randomly from a Rayleigh distribution \cite{Holthuijsen-2010}. 
The mean value of the Rayleigh distribution is $\sqrt{2 S(\omega_n) \Delta\omega}$, where $S(\omega)$ is the one-dimensional energy density spectrum and $\Delta \omega$ is the uniform frequency spacing. 

For a given spectrum, $S$, the free surface of the water at a horizontal location $x$ and time $t$ for the incoming waves is
\begin{equation}
\eta_I(x,t) = \sum_{n=1}^{N} A_{n} e^{i(k_n x - \omega_n t + \theta_n)}.
\label{eqn:IrregularWave_Components}
\end{equation}
The wavenumber $k_n$ satisfies the linear, finite-depth dispersion relation, 
with a depth of 1000\,m that approximates deep water conditions, i.e.\
$k_{n}=\omega_{n}^{2}/g$, where $g\approx{}9.81$\,m\,s$^{-2}$ is the acceleration due to gravity.

The spectrum used for the results in \S\S\ref{sec:Model_Comp}--\ref{sec:predictions} is the JONSWAP spectrum \cite{Hasselmann-1973}, defined by
\begin{subequations}
\label{eqn:Spectra_JS}
\begin{equation}
S_{jp}(\omega;\tau_p) = 2\pi \beta g^2 \omega^{-5} \exp{\left(-\frac{5}{4} \left(\frac{\omega}{\omega_p} \right)^{-4} \right) } \gamma^{\Lambda}
\end{equation}
where
\begin{equation}
\gamma = 3.3, \quad
\Lambda=
 \exp\left(-\dfrac{\left(\frac{\omega}{\omega_p} - 1\right)^2}{2 \left(\sigma(\frac{\omega}{\omega_p} )\right)^2}\right)
 \quad\text{and}\quad
 \sigma\left(x\right) = \begin{cases} 
 0.07 & x < 1 \\
 0.09 & \text{otherwise} .\\
 \end{cases}
\end{equation}
\end{subequations}
The spectrum is parameterised by a peak period $\tau_p$ (peak frequency $\omega_p = 2 \pi /\tau_p$) where it attains its maximum value. The spectra is also modified by the ``Phillips'' parameter $\beta$. 

The numerical results in \S\S\ref{sec:validation}--\ref{sec:predictions} are used to investigate how the significant wave height, $H_s$, and the peak period, $\tau_p$, affect overwash extent. 
To facilitate the investigation, the JONSWAP spectrum \eqref{eqn:Spectra_JS} is normalised to
\begin{equation}
\Tilde{S}_{jp}(\omega;\tau_p,H_s) = \frac{H_s^2}{16} \dfrac{S_{jp}(\omega;\tau_p) }{ \int_0^\infty S_{jp}(\omega;\tau_p) d\omega },
\label{eqn:JS_Norm}
\end{equation} 
which fixes the spectral shape given by $\gamma$ and $\sigma$ and replaces the scale dependence on $\beta$ with $H_s$. 

\subsubsection{Coupled Floe--Wave Motions}
\label{subsec:Floe--Wave}
A two-dimensional coupled floe--wave motion model for incoming irregular waves is produced by linear superposition of a regular wave (single frequency) version of the model. 
The regular wave model is described first and then extended to the irregular wave model. The regular wave model is subsequently regarded as a degenerate case of the irregular wave model. 

\begin{figure}
\centering
\includegraphics[width=0.7\textwidth]{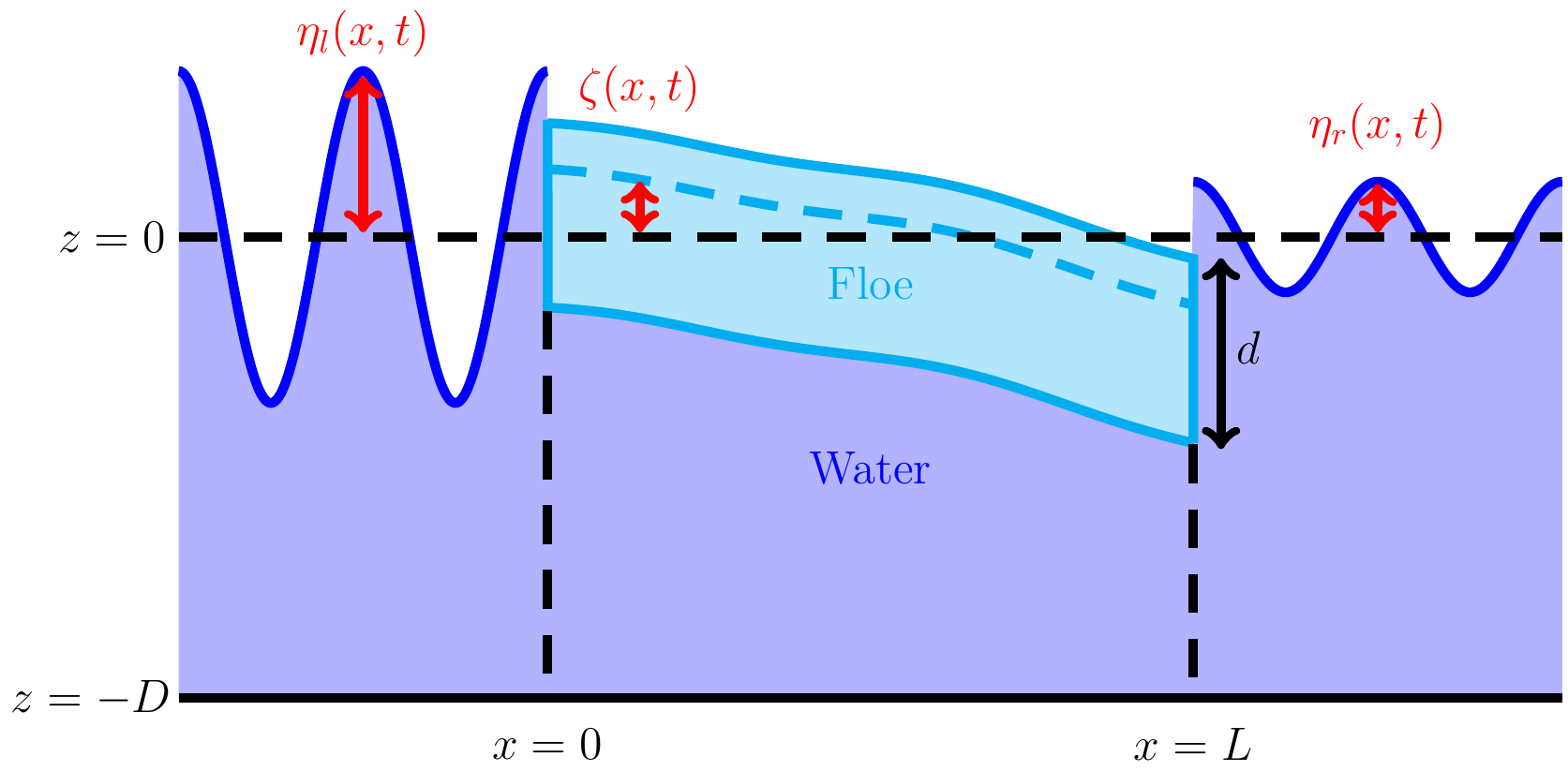}
\caption{Example of coupled floe--wave motion solution~\eqref{eqn:Coup_WF_Sol}, for a floe of length $L$, and thickness $d$ and a regular incoming wave with amplitude $A$ and frequency $\omega$.}
\label{Fig:Model}
\end{figure}

The horizontal axis ($x$) has its origin at the left edge of the floe, and the vertical axis ($z$) has its origin at the still water line (Figure~\ref{Fig:Model}). 
The water has constant density $\rho$, where, unless otherwise stated, $\rho = 1025 \,\text{kg} \,\text{m}^{-3}$ and finite-depth water that approximates the deep water conditions (depth is $1000$\,m).

The floe has length $L$, thickness $d$ and density $\rho'$, and, hence, Archimedean draught $d_d = (\rho'/\rho) \, d$ and freeboard $d_f = d - d_d$. The elastic response of the floe is governed by a Young's modulus $E$, which controls the stiffness, and a Poisson ratio $\nu$, which describes expansion perpendicular to the direction in which force is applied. 
The standard values for sea ice, $E = 6\,\text{GPa}$, $\nu= 0.3$ and $\rho' = 920 \,\text{kg} \,\text{m}^{-3}$, are used \cite{Timco-2010,Bennetts-2012}. 

Linear potential flow theory governs the motion of the water surrounding and under the floe. The water surrounding the floe has free surface $z=\eta(x,t)$ ($x<0$ and $x>L$). 
Kirchoff--Love thin plate theory is used to model the motion of the floe in terms of the vertical displacement of its neutral plane, $\zeta(x,t)$ ($0<x<L$), so that the bottom of the floe is at $z= \zeta(x,t)- d_d $ and the top of the floe is $z=d_f + \zeta(x,t)$. 
The floe displacement is a combination of rigid-body motions, heave and pitch, and elastic motions \cite{Montiel-2013b}. 

Figure~\ref{Fig:Model} shows a snapshot of the floe displacement and surrounding wave field in response to an incoming regular wave (from the left) with angular frequency $\omega$ and amplitude $A$, solved using the variational method of \citeA{Bennetts-2007}. The wave field to the left of the floe, $\eta(x,t)\equiv\eta_{l}(x,t)$ ($x<0$), is a combination of the incoming regular wave and a regular reflected wave, while the wave field on the right, $\eta(x,t)\equiv\eta_{r}(x,t)$ ($x>L$), is the transmitted regular wave. 
The solutions are written in the form
\begin{subequations}\label{eqn:Coup_WF_Sol}
 \begin{align}
 \eta_l(x,t) &= A \,\Re\left\lbrace \left(e^{i k x } +R(\omega)e^{-i k x} \right) e^{-i \omega t } \right\rbrace\\
 \eta_r(x,t) &= A \, \Re\left\lbrace T(\omega)e^{i k(x - d)} e^{-i \omega t } \right\rbrace \\
 \zeta(x,t) &= A \, \Re\left\lbrace \hat{\zeta}(x,\omega) e^{-i \omega t }\right\rbrace,
 \end{align}
\end{subequations} 
where $R(\omega)$ and $T(\omega)$ are complex-valued reflection and transmission coefficients, respectively, and $\hat{\zeta}(x,\omega)$ is the complex-valued floe displacement profile. 
Evanescent modes, which decay exponentially away from the floe edges, are included in the calculations of $R(\omega)$, $T(\omega)$ and $\hat{\zeta}$, but are neglected in expressions (\ref{eqn:Coup_WF_Sol}a--b), following \citeA{Skene-2015}.

{
The reflection and transmission coefficients and the displacement profile at the floe edges depend on the incoming $\omega$, the floe thickness $d$ and the floe length $L$. Figure~\ref{Fig:Depend} shows these dependencies for the absolute value of these coefficients for a pancake floe with $d=0.5\,\text{m}$ and $L=0.7\,\text{m}$. In Figure~\ref{Fig:Depend}(a) the floe geometry is fixed and the $\omega$ of incoming waves is varied. For Figures~\ref{Fig:Depend}(b--c) the period is fixed to $\tau = 2\pi / \omega = 10\,\text{s}$, which is comparable to typical $\tau_p$ in the Southern Ocean \cite{Young-2020}. In Figure~\ref{Fig:Depend}(b) the length is fixed and thickness ($0.7$ shown as grey dashed line) is varied and in Figure~\ref{Fig:Depend}(c) the thickness is fixed and the length ($0.5$ shown as grey dashed line) is varied. Extreme floe geometries are included in Figures~\ref{Fig:Depend}(b--c) to demonstrate large scale trends qualitatively representative of the trends experienced for realistic floe geometries. 


The dependence of the wave--floe motion coefficients in Figures~\ref{Fig:Depend}(a--c) can be described by the relative size of the incoming wavelength $\lambda= 2 \pi / k$ to the floe thickness and floe length \cite{Meylan-1994,Bennetts-2012a}. The dependence is produced by the flexural-gravity wave induced in the coupled floe--water motions, as explained by \citeA{Meylan-1994}. The system conserves energy, so that $|R(\omega)|^2 + |T(\omega)|^2 =1$, and, hence, an increase in reflected wave energy comes at the expense of transmitted wave energy and vice versa. When the wavelength is short compared to the floe thickness and length (high frequencies), then most of the incoming wave energy is reflected and the floe is essentially still, so that $|R(\omega)| \approx 1$ and $\vert\hat{\zeta}(x,\omega)\vert \approx 0$ with conservation ensuring that $|T(\omega)| \approx 0$. When the wavelength is long compared to the floe thickness and length (low frequencies), then most incoming wave energy is transmitted and the floe moves in-phase with the waves, so that $|T(\omega)| \approx 1$ and $\vert{}\hat{\zeta}(x,\omega)\vert \approx 1$ with conservation ensuring that $|R(\omega)| \approx 0$. The black dotted lines in Figure~\ref{Fig:Depend}(a) indicate the short- and long-wave regimes to within a $1\%$ tolerance of their asymptotic values. For the chosen floe, the long-wave regime occurs when $\omega < 0.7 \, \text{s}^{-1}$ and $ \lambda 	> 162 d = 116L$, while the short-wave regime occurs when $\omega > 7.4 \, \text{s}^{-1}$ and $ \lambda < 1.5d = 1.1L $. The coupled floe--wave motion transitions between these two regimes as $\omega$ and $d$ increase, with some resonances apparent \cite{Meylan-1994,Lever-1988}. Increasing floe length also increases reflection, particularly when $L$ is small. As $L$ increases further there is no overall trend as resonance effects dominate, producing consistent oscillations in coefficient values for large floe lengths in Figure~\ref{Fig:Depend}(b). It is demonstrated in \S\ref{sec:validation} that resonance effects average out for irregular waves leaving only the overall trends with respect to long and short wavelengths. 

\begin{figure}
\centering
\includegraphics[width=\textwidth]{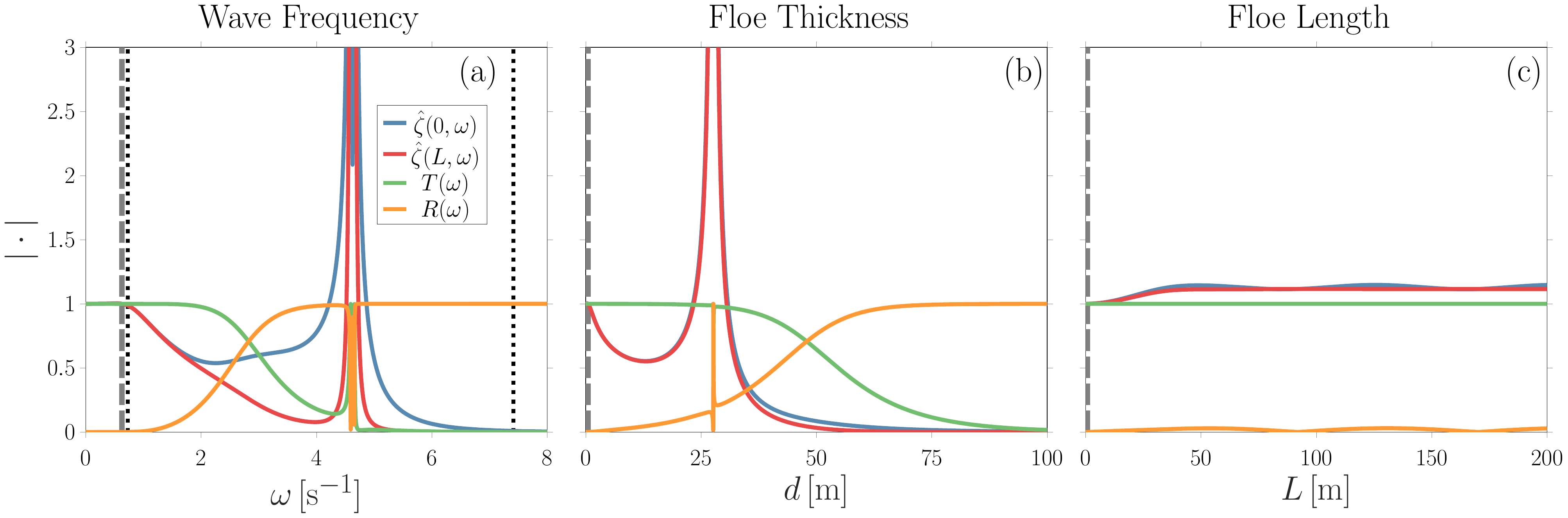}
\caption{Dependence of the absolute value of transmission, reflection and plate edge coefficients in \eqref{eqn:Coup_WF_Sol} on (a)~wave frequency $\omega$, (b)~floe thickness $d$ and (c)~floe length $L$. The fixed values (a)~$\tau = 10 \,s$, (b)~$d=0.7\,\text{m}$ and (c)~$L=0.5\,\text{m}$ are highlighted ({\color{gray}\dashedrule}). In~(a) the short- and long-wave limits are indicated ({\dotrule}). }
\label{Fig:Depend}
\end{figure}
}

The coupled floe--wave motions for an irregular incident wave are calculated as the linear superposition for the regular wave components, $\omega_1,\dots,\omega_N$, similar to \eqref{eqn:IrregularWave_Components}. 
For each $\omega_n$, the reflection, transmission and floe displacement profile coefficients are calculated, as described above.
Since the problem is linear, for an incoming wave spectrum $S$, the wave spectra of the left ($\eta_l$) and right ($\eta_r$) of the floe at $\omega_n$ \cite{Holthuijsen-2010} are
\begin{equation}
 S_{\eta,l}(\omega_n) = \left|1 + R(\omega_n)\right|^2 S(\omega_n) , \quad S_{\eta,r}(\omega_n) = \left| T(\omega_n)\right|^2 S(\omega_n)
 \label{eqn:Spec_Motion}
\end{equation}
and the spectrum of floe displacement profile ($\zeta$) is
\begin{equation}
 S_{\zeta}(\omega_n,x) = \left| \hat{\zeta}(x,\omega_n)\right|^2 S(\omega_n).
 \label{eqn:Spec_MotionFloe}
\end{equation}
The spectra (\ref{eqn:Spec_Motion}--\ref{eqn:Spec_MotionFloe}) act as inputs to the overwash of a single floe model.

\subsubsection{Overwash of a Single Floe}\label{sec:onset_irregular}
Overwash is generated when the either floe edge moves below the surrounding water surface. 
Thus, an overwash event at time $t$ occurs if either $\psi_l(t) = \eta_l(0,t) - \zeta(0,t) $ or $\psi_r(t) = \eta_r(d,t) - \zeta(L,t)$ are greater than the equilibrium freeboard, $d_{f}$.
From \eqref{eqn:Spec_Motion} and \eqref{eqn:Spec_MotionFloe}, the spectra describing $\psi_{l}$ and $\psi_{r}$ for each $\omega_n$ are, respectively,
\begin{align}
S_l(\omega_n) = \left|1 +R(\omega_n) - \hat{\zeta}(0,\omega_n) \right|^2 S(\omega_n) 
\quad{}\text{and}\quad{}
S_r(\omega_n) = \left|T(\omega_n) - \hat{\zeta}(d,\omega_n) \right|^2 S(\omega_n).
 \label{eqn:SpectralResponse}
\end{align}
The spectra define ensembles of possible time series, where each realisation is given by sampling the associated distributions of phases and amplitudes of the regular wave components. From the respective edge spectra \eqref{eqn:SpectralResponse},
the ensemble average of the time between overwash events
is \cite{Rice-1945}
\begin{equation}
 \overline{\tau}_{E}(d_f) =
 \exp\left(\dfrac{d_f^2}{2m_0 } \right)
 \sqrt{\dfrac{m_0}{m_2}} 
 \quad\text{where}\quad
 m_j = \int_{0}^{\infty} \omega^j S_{E}(\omega^j) d\omega
 \quad (j=0,2),
 \label{eqn:Mean_Per}
\end{equation}
for the left edge ($E = l$) and the right edge ($E = r$). The spectral moments $m_j$ are calculated by quadrature using $\omega_1,\dots,\omega_N$ as the quadrature points. Additionally, $\omega_1$ and $\omega_N$ are chosen so that $S(\omega_1)$ and $S(\omega_N)$ are very small, to ensure that the integrals are well approximated. A further height tolerance of $\epsilon$ is added, so that the mean time between overwash events at the left and right edges is $\overline{\tau}_l(d_f + \epsilon)$ and $\overline{\tau}_r(d_f + \epsilon)$, respectively. The value $\epsilon= 0.001 \,\text{m}$ is used in the results presented in \S\ref{sec:validation} and \S\ref{sec:predictions}.

To provide a consistent measure of the prevalence of overwash events for a variety of incoming waves, the relative frequency of overwash events
\begin{equation}
 f_o = \max \left\lbrace \dfrac{\overline{\tau}_l(d_f+\epsilon) }{\overline{\tau}_I(0)} , \dfrac{\overline{\tau}_r(d_f+\epsilon) }{\overline{\tau}_I(0)} \right\rbrace,
 \label{eqn:OW_Freq_1}
\end{equation}
is defined, where $\overline{\tau}_I(0)$ is the mean wave period for the incoming spectrum $S$ calculated using \eqref{eqn:Mean_Per}. The relative frequency of overwash events is the maximum of the relative overwash event frequency at the left and right edges of the floe, since overwash events can occur at either edge.
The floe is judged to experience overwash using \eqref{eqn:OW_Freq_1} if $f_o > f_{tol}$ where $f_{tol}$ is a chosen threshold for relative frequency of overwash events.
In \S\ref{sec:validation} and \S\ref{sec:predictions}, $f_{tol} = 0.05$ is used, so that a floe is determined to be overwashed when at least one overwash event occurs over $20$ mean periods.

For regular waves
\begin{equation*}
 \psi_l(t) = 
 A Re\left\lbrace (1 +R(\omega) - \hat{\zeta(0,\omega)}) e^{-i \omega t } \right\rbrace \quad \text{and}\quad \psi_r(t) = A Re\left\lbrace (T(\omega) - \hat{\zeta(L,\omega)}) e^{-i \omega t } \right\rbrace.
\end{equation*}
Therefore, the mean time between overwash events (waves above $d_f + \epsilon$ at the floe edges) is
\begin{align*}
 \overline{\tau}_l(d_f + \epsilon) &=\begin{cases}
 \tau &\text{if } A |1 + R(\omega) - \hat{\zeta}(0,
 \omega)| > d_f + \epsilon \\
 0 & \text{otherwise}
 \end{cases}\\
 \text{and}\quad
 \overline{\tau}_r(d_f + \epsilon) &= \begin{cases}
 \tau &\text{if } A |T(\omega) - \hat{\zeta}(L,
 \omega)| > d_f + \epsilon \\
 0 & \text{otherwise},
 \end{cases}
\end{align*}
at the left and right edges respectively. Consequently, the relative frequency of overwash events for regular waves is $f_o = 1$ when $A |1 +R(\omega) - \hat{\zeta}(0,\omega)| > d_f + \epsilon$ or $A |T(\omega) - \hat{\zeta}(L,\omega)| > d_f + \epsilon$, otherwise $f_o = 0$.

\begin{figure}
\centering
\includegraphics[width=0.8\textwidth]{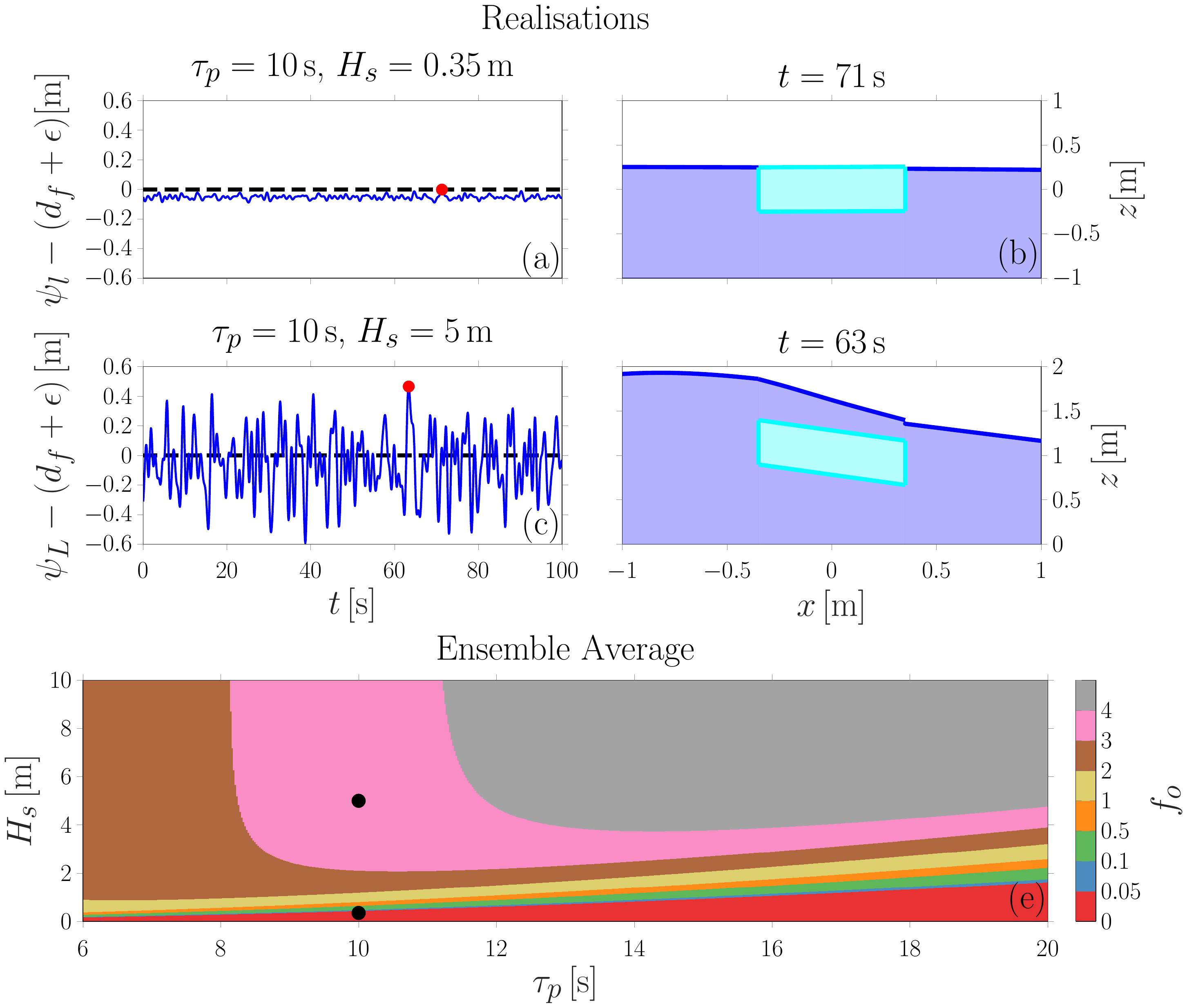} 
\caption{Deterministic simulations of realisations of the spectrum compared to ensemble averages: (a) and (c) time series of the left wave field relative to top of the left edge of floe for non-overwashing and overwashing examples respectively (\ColorCircle{black} in (e)), (b) and (d) simulation for non-overwashing and overwashing examples respectively produced at highlighted time (\ColorCircle{red} in (a) and (c)) and (e) ensemble average of overwash event frequency for a variety of incoming wave spectra.}
 \label{Fig:OW_Q_Ex}
\end{figure}

Figures~\ref{Fig:OW_Q_Ex}(a,c) show example realisations (process described in \ref{App:Sec_SWE}) of 
$\psi_{l}-(d_f + \epsilon)$, i.e.\ 
the free surface of the water relative to the top of the floe at the left edge of the floe, over arbitrarily chosen $100$\,s intervals. The incoming waves are realisations of spectrum \eqref{eqn:JS_Norm} with $\tau_p = 10 \,\text{s}$
and (a)~$H_{s}=0.35$\,m as an example of no overwash being predicted, and (c)~$H_{s}=5$\,m, as a mean $H_s$ in Southern Ocean \cite{Young-2020}.
The floe is a pancake floe, which has length $L = 0.7\, \text{m}$ and thickness $d =0.5\, \text{m}$.
Figures~\ref{Fig:OW_Q_Ex}(b,d) show corresponding snapshots from the coupled floe--wave model at the time when $\psi_l - (d_f + \epsilon)$ is at its maximum over the $100$\,s interval (red circles), which are extended to model overwash on the floe surface using the shallow water wave equation (\ref{App:Sec_SWE}), in a similar manner to \citeA{Skene-2015}.

The realisation with $H_s = 0.35\, \text{m}$ (Figure~\ref{Fig:OW_Q_Ex}a) has no overwash events ($\psi_l - (d_f + \epsilon) < 0$), as shown in the snapshot for maximum $\psi_{l}-(d_f + \epsilon)$ (Figure~\ref{Fig:OW_Q_Ex}b). Thus, for the realisation with $H_s = 0.35\, \text{m}$, the relative overwash event frequency is zero. Overwash events occur frequently for the realisation with $H_s = 5\, \text{m}$ (Figure~\ref{Fig:OW_Q_Ex}c), with a relative overwash event frequency for the realisation of $5.3$, i.e.\ almost five overwash events per mean wave period. The snapshot at the maximum (Figure~\ref{Fig:OW_Q_Ex}d) shows overwash being forced at the left edge and right edge of the floe, completely submerging the floe. While overwash is generated at the right edge with an overwash event frequency of $3.5$, more water is pushed onto the floe at the left edge so that it is the largest source of overwash.

Figure~\ref{Fig:OW_Q_Ex}(e) shows the ensemble average of relative overwash event frequencies, $f_{o}$, for the floe used in the realisations ($L = 0.7\, \text{m}$ and $d = 0.5\, \text{m}$) over a range of peak periods and significant wave heights. The $H_s$--$\tau_p$ combinations for the realisations in Figures~\ref{Fig:OW_Q_Ex}(a--d) are indicated (black circles). 
For the incoming spectrum used for Figures~\ref{Fig:OW_Q_Ex}(a,b) ($\tau_p = 10\, \text{m}$ and $H_s = 0.35\, \text{m}$), 
the ensemble average is $f_{o}= 0.003 \ll f_{tol}=0.05$. Therefore, overwash events are rare
for the ensemble of incoming waves described by this spectrum. The ensemble average of relative overwash event frequency, $f_o$, compares well to the realisation in which the relative overwash event frequency is zero (Figure~\ref{Fig:OW_Q_Ex}a).
In contrast, the incoming spectra with $\tau_p = 10\, \text{s}$ and $H_s = 5\, \text{m}$ has $f_o=3.5\gg{}f_{tol}$, so there are approximately four overwash events every mean period on average. The relative overwash event frequency of the ensemble is comparable to the relative overwash event frequency of $5.3$ for the realisation in Figures~\ref{Fig:OW_Q_Ex}(c,d), in which overwash events are frequent. 
Although the dependence of the wave and floe displacement coefficients on $\omega$ indicates the presence of resonance effects in Figure~\ref{Fig:Model}, the $f_o$ contours are smooth and there are no anomalous behaviours for particular $\tau_p$. Thus, resonance effects are not apparent in the ensemble average overwash predictions.

\subsection{Overwash Extent of Floe Fields}

\subsubsection{Floe Field Properties}
\label{subsubsec:FSD}
A field of floes is assumed to be composed of separated free-floating floes with no floe--floe interactions. The floe field is defined by a constant ice thickness and floe concentration, and a prescribed FSD.
The ice thickness and floe concentration are chosen to be constant in a given floe field for simplicity, rather than due to a limitation of the model.
In \S\ref{subsec:Agulhas}, the constant floe concentration restriction is relaxed to match floe concentration variations in an observed MIZ. 

The FSD is defined by the exceedance probability distribution $P^*(L)$, which gives the probability of finding a floe greater than length $L$ in the floe field. 
A split power law FSD is used, based on FSD observations in the MIZ \cite{Toyota-2011,Alberello-2021}. The exceedance probability given by a split power law is
\begin{subequations}\label{eqn:FSD_power}
 \begin{equation}
 P^*(L) = \begin{cases} 
 1 & \text{if} \quad L < L_{min}\\
 \left(1 - \alpha \right) \beta_1 L^{-\gamma_1} & \text{if} \quad L_{min} \le L \le L_{crit} \\
 \alpha \beta_2 L^{-\gamma_2} & \text{if} \quad L_{crit} < L,
 \end{cases}
\end{equation}
where the parameters
\begin{equation}
 \beta_1 = \frac{1}{L_{min}^{-\gamma_1} - L_{crit}^{-\gamma_1}} , \quad \beta_2 = \frac{1}{L_{crit}^{-\gamma_2}} \quad\text{and} \quad \alpha =\left( 1 + \dfrac{\beta_2\gamma_2 L_{crit}^{-\gamma_2-1}}{ \beta_1\gamma_1 L_{crit}^{-\gamma_1-1} } \right)^{-1}
\end{equation}
\end{subequations}
ensure $P^*(L_{min}) = 1$ and that the probability distribution is continuous at $L_{crit}$ where the two power laws intersect. The parameters $\gamma_1$, $\gamma_2$ and $L_{crit}$ are obtained by fitting floe size data, while $L_{min}$ is the smallest measurable floe size. 

For computations, a finite number of floe lengths, $L_1,\dots, L_{M}$, are used. The floe lengths are uniformly spaced, i.e.\ $\Delta L = L_{m+1} - L_{m}$ is fixed for all $m=1,\ldots,M-1$. The exceedance probability $P^*$ produces a probability distribution
\begin{equation}
\label{eqn:Discrete_FSD}
p(L_m) = P^*(L_m - \Delta L/2) - P^*(L_m + \Delta L/2)
\end{equation}
for the finite number of floe lengths, called the discrete FSD. This probability distribution results in all floes of length $L_m - \Delta L/2 \le L \le L_m + \Delta L/2$ in the continuous FSD being represented by a floe of length $L_m$ in the discrete FSD. 

The discrete FSD represents an ensemble of possible floe fields where each realisation is generated by repeated sampling of the probability distribution \eqref{eqn:Discrete_FSD}. The resultant realisation of the floe field will have a number of floes of length $L_m$, $N_f(L_m)$ and the proportion of $N_f(L_m)$ to the total number of floes will approach $p(L_m)$ as the total number of floes increases. A field of floes with a fixed floe concentration $c_f$ over a distance $x$ will cover a distance of $c_fx$ when placed end to end. For a floe field described by the FSD \eqref{eqn:Discrete_FSD}, the expected number of floes covering a total end to end distance of $c_fx$ is $c_fx / \overline{L}$ where $\overline{L}$ is the mean floe size in the FSD \eqref{eqn:Discrete_FSD}. 

A pancake floe field and a floe field comprised of a fragmented ice cover will be investigated in \S\ref{sec:predictions}. 
The FSD of a pancake floe field in an Antarctic MIZ was measured by \citeA{Alberello-2019}, who found the FSD values $\gamma_1=1.1$, $\gamma_2 = 9.4$, $L_{crit} = 3.15\, \text{m}$ and $L_{min} = 0.25 \, \text{m}$ for \eqref{eqn:FSD_power}. The sea ice floe concentration was $c_f\approx{}0.6$ throughout the measurement period. The thickness of pancake ice was not measured by \citeA{Alberello-2019}, but is typically in the range of $0.1\text{--}1\,\text{m}$ \cite{Worby-1996,Wadhams-2018} and thus $d = 0.5\, \text{m}$ will be taken as the constant thickness for a pancake floe field. The floe field generated by a fragmented ice cover in the Weddell Sea was measured by \citeA{Toyota-2011}, who found FSD values of $\gamma_1=1.39$, $\gamma_2 = 5.18$ , $L_{crit} = 30\, \text{m}$ and $L_{min} = 2 \, \text{m}$ for \eqref{eqn:FSD_power}. The ice concentration varied between $c_i=0.3\text{--}1$, and, therefore, it is reasonable to use $c_f=0.6$ for the fragmented floe field making it consistent with the pancake floe field. The $95\%$ confidence interval for the thickness of the floes was measured as $ 1.08 \pm 1.07 \, \text{m}$, the mean thickness $d = 1.08$\,m will be taken as the constant thickness for fragmented floe fields.

\subsubsection{Wave Attenuation in the MIZ}\label{sec:wave_atten}
Wave attenuation over distance travelled through the MIZ is modelled using the function $\mathcal{T}(\omega,x)$, defined by
\begin{equation}
 S(\omega,x) = \mathcal{T}(\omega,x) S(\omega,0),
 \label{eqn:Attn_Model}
\end{equation}
where $x$ is the distance into the MIZ from the open ocean, and $S(\omega,0)$ is the incoming spectrum for the MIZ. 
The attenuation function is expressed as the product
\begin{equation*}
 \mathcal{T}(\omega,x) = T_{scat}(\omega,x) T_{diss}(\omega,x) ,
 \label{eqn:Trans_field}
\end{equation*}
where $\mathcal{T}_{scat}(\omega,x)$ represents wave attenuation due to scattering and $\mathcal{T}_{diss}(\omega,x)$ represents attenuation due to dissipative effects. Attenuation due to scattering is modelled using the coupled floe--wave motions of all floes in the floe field and thus depends on the FSD. In contrast, an empirical model is used for dissipation \cite{Meylan-2014}, which only depends on $\omega$ and $x$ and not the FSD. 

Attenuation due to scattering varies according to the realisation of the discrete FSD. 
To produce $\mathcal{T}_{scat}(\omega,x)$, the ensemble average of attenuation due to all realisations of the discrete FSD with a floe concentration $c_f$ over a distance $x$ is used. An individual floe of length $L_m$ transmits a spectrum, which is attenuated by $\left| T(\omega,L_m) \right|^2$ due to scattering. Given a floe concentration $c_f$, a floe of length $L_m$ occurs on average $ q=p(L_m) c_f x / \overline{L}$ times over the distance $x$, where $\overline{L}$ is the average floe length for the FSD. The average attenuation in the ensemble described by the discrete FSD is then
\begin{equation}
 \mathcal{T}_{scat}(\omega,x) = \prod_{m=1}^{M}\left(\left| T(\omega,d_m) \right|^2\right)^{q}.
 \label{eqn:FSD_TransMean}
\end{equation}
For a deterministic floe field (as in \S\ref{sec:validation}), the number of floes and floe lengths over the distance $x$ are known. When the floe field is deterministic, the known counts of the respective floe lengths replace the exponent $q$. 

\citeA{Meylan-2014} derived an empirical model for attenuation over distance, based on measurements in an Antarctic MIZ. The relatively large wavelengths to floe lengths in the study area suggest that scattering was negligible and, hence, dissipation was the dominant source of attenuation \cite{Squire-2020}. The empirical model has been found to agree with measurements of attenuation in different wave and ice conditions \cite{Meylan-2018a}. The empirical dissipation attenuation model is
\begin{equation}
 \mathcal{T}_{disp}(\omega,x) = e^{- \left(a_1 \left(\frac{\omega}{2\pi} \right)^{2} +a_2 \left(\frac{\omega}{2\pi} \right)^{4} \right) x},
 \label{eqn:Trans_long}
\end{equation}
with $a_1 = 2.12 \times 10^{-3} \, \text{s}^2 \, \text{m}^{-1}$ and $a_2 = 4.59 \times 10^{-2} \, \text{s}^4 \, \text{m}^{-1}$. Proposed dependence of $\mathcal{T}_{diss}$ on ice concentration \cite{Bennetts-2017} has not yet received supporting evidence, and, thus, is omitted from the model for simplicity. 

\begin{figure}
\centering
\includegraphics[width=\textwidth]{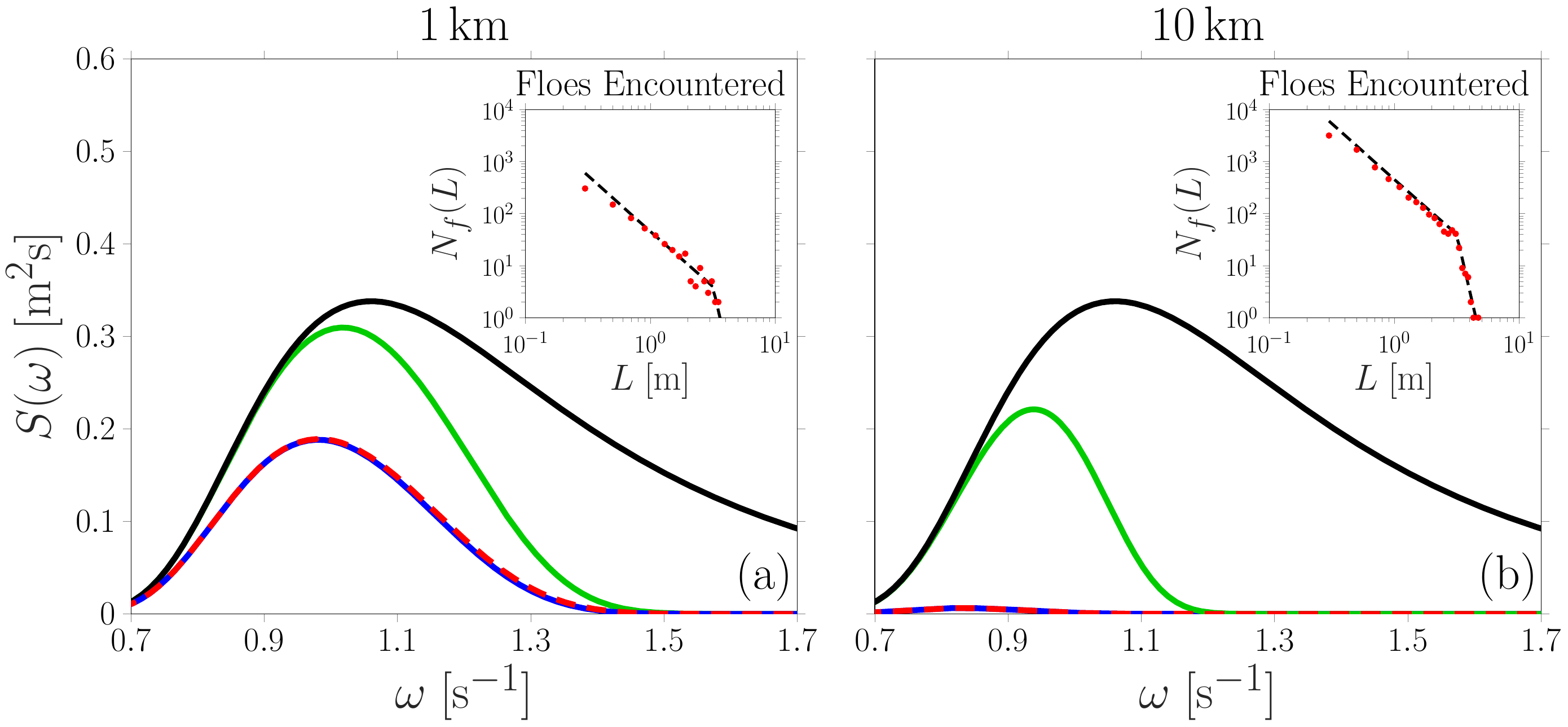}
\caption{Attenuation of an example JONSWAP spectrum (\solidrule[2mm]) through the pancake FSD with the ensemble average spectrum with ({\color{blue}\solidrule[2mm]}) and without ({\color{green!80!black}\solidrule[2mm]}) $\mathcal{T}_{disp}$ and the attenuated spectrum after passing through a realisation of the discrete FSD ({\color{red}\dashedrule}) over distances of (a)~$1\,\text{km}$ and (b)~$10\,\text{km}$. The inset shows the respective count of the number of floes encountered $N_f(L)$ for each floe length $L$ in the sample from the discrete FSD (\ColorCircle{red}) (which causes the attenuated spectrum ({\color{red}\dashedrule}) ) compared to the underlying discrete FSD multiplied by expected number of floes ($c_f x/ \overline{L}$) (\dashedrule). }
\label{Fig:OW_Q_Trans}
\end{figure}

Figure~\ref{Fig:OW_Q_Trans} shows the attenuation of an incoming spectrum \eqref{eqn:JS_Norm} with $H_s = 2 \, \text{m}$ and $T_p = 6 \, \text{s}$ (from relationship \eqref{eqn:Energy_rel} in \S \ref{subsubsec:EffIncWave})
at (a)~$x=1$\,km and (b)~$10$\,km into
a pancake floe field. 
Four spectra are displayed in each panel: the initial spectrum (black); the ensemble average spectrum for the FSD with only attenuation due to scattering ($\mathcal{T}_{diss} = 1$; green); 
the ensemble average spectrum for the FSD with attenuation due to scattering and dissipation (blue); and a spectrum attenuated by a realisation of the FSD (red dashed), where the insets show the counts for each floe in the realisation. 

Scattering dominates attenuation when wavelengths are comparable to floe lengths \cite{Squire-1995,Bennetts-2015}, which is the high-frequency (short-period) regime $\omega > 1.4\,\text{s}^{-1}$ for the pancake floe field. 
However, $\omega_p = 0.79\,\text{s}^{-1}$, is below the high-frequency regime and so dissipation dominates attenuation for this spectrum. 
The attenuation reduces $H_s$ by $24\%$ after $1\,\text{km}$ and by $38\%$ after $10\,\text{km}$. The reduction in significant wave height produced by the attenuation model \eqref{eqn:Attn_Model} is greater than the $12\%$ reduction over roughly $15\, \text{km}$ observed by \citeA{Alberello-2021}, which is attributed to evidence of wave generation by winds in the MIZ and the changing wave conditions during those measurements. 

The preferential attenuation of higher frequency waves results in a gradual increase in peak period, $\tau_p$, as the spectrum propagates through the floe field. The ensemble average attenuated spectrum \eqref{eqn:FSD_TransMean} and the attenuated spectrum of the realisation are almost identical due to the dominance of $\mathcal{T}_{diss}$, as well as the proximity of the floe counts in the realisation to the underlying probability distribution. 
The realisation and the ensemble average attenuated spectra do not demonstrate resonance effects from the coupled--floe wave motions, as they are smooth and lack the sudden peaks seen in Figure~\ref{Fig:Depend}.

\subsubsection{Overwash Extent}
The ensemble of floe fields described by the FSD is used in combination with the attenuation function \eqref{eqn:Attn_Model} to produce an ensemble average attenuated spectrum at some distance $x$ into the floe field $S(\omega,x)$. The ensemble of irregular waves described by the attenuated spectrum $S(\omega,x)$ is then used to determine the ensemble average overwash event frequency for a given floe at that distance $x$ into the floe field. To quantify the extent of overwash into a floe field, two measures of overwash extent are used.

The first measure is used to predict the overwash extent for a floe of chosen thickness $d$ and length $L$. 
For floes with these particular properties, the relative overwash event frequency at $x$ 
is $f_o(x;d,L)$, i.e.\ previous implicit dependencies have been made explicit. 
The maximum extent for a floe with thickness $d$ and length $L$, $X_{d,L}$, is the largest distance $x$ such that
\begin{equation*}
 f_o(x;d,L) > f_{tol},
\end{equation*}
i.e.\ the largest distance the floe with known properties is predicted to be overwashed by the single floe overwash model. 

The second measure is used to predict the expected overwash extent for all floes in the FSD at a set distance $x$. Thus, it uses the expected relative frequency of overwash events for all floes at $x$, which is
\begin{equation*}
 \overline{f}_o(x;d) = \sum_{m=1}^{M} f_o(x;L_m,d) p(L_m).
\end{equation*}
It provides an indication of the relative frequency of overwash for most floes in the floe field described by the FSD. 
The average extent $\overline{X}_d$ is the largest distance $x$ such that
\begin{equation*}
 \overline{f}_o(x;d) > f_{tol}.
\end{equation*}
It is the largest distance for which most floes in the floe field are predicted to be overwashed by the single floe overwash model.

\section{Comparison with Wave Basin Experiments}
\label{sec:validation}
\begin{figure}
\centering
\includegraphics[width=\textwidth]{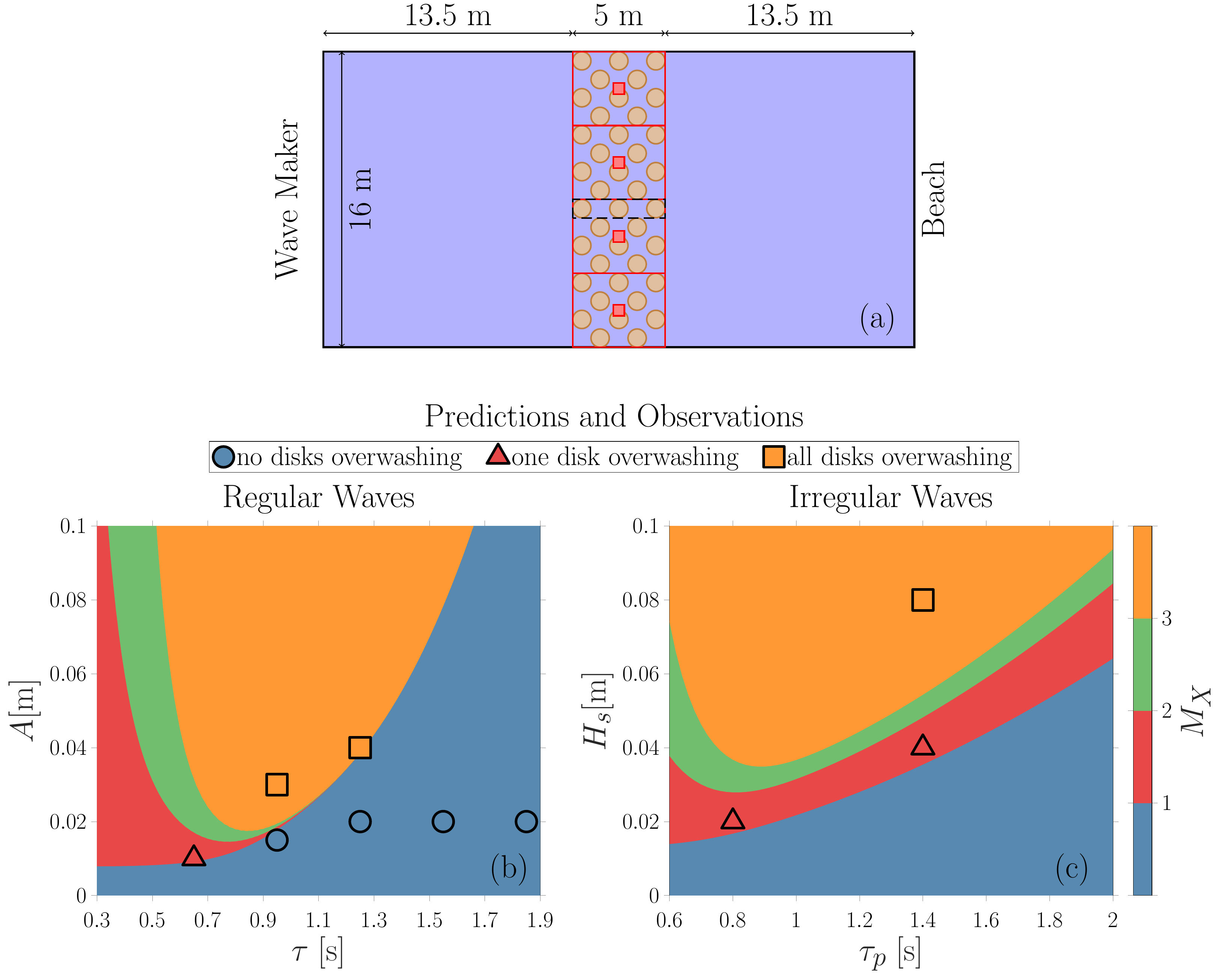}
 \caption{(a)~ Schematic of the experimental set-up showing wooden disks (\BoundCircle{solid}{1pt}{brown}{brown!50}) with three disk transect highlighted (\BoundSquare{dashed}{1pt}{black}{white}), camera locations (\BoundSquare{solid}{1pt}{red}{red!50}) with camera field of view (\BoundSquare{solid}{1pt}{red}{white}). Comparison of model predictions of number of disks overwashed ($M_X$) with experimental results for the (b)~regular and (c)~irregular wave experiments. The colour bar describes the model result by highlighting the bounds of $M_X$, so \ColorSquare{PRSA_1} corresponds to no floes overwashing (less than $1$) while \ColorSquare{PRSA_2} corresponds to one floe overwashing (above $1$ but less than $2$). }
 \label{Fig:OW_PRSA}
\end{figure}

A laboratory experiment that investigated interactions between incoming regular and irregular waves and an array of model floes (wooden disks) was conducted in Basin de G\'{e}nie Oc\'{e}anique FIRST wave basin facility, located at Oc\'{e}anide, La Seyne sur Mer, France. A schematic of the experimental set-up is given in Figure~\ref{Fig:OW_PRSA}(a). In the laboratory experiment, the wave maker on the left of the tank generates rightwards propagating waves, which are transmitted and reflected as they pass through the array of disks (brown circles). The transmitted waves are then dissipated by the beach. Four cameras (red squares) are situated above the disks, allowing them to be observed throughout the experiment. 

The disks have density $\rho' = 545 \, \text{kg} \,\text{m}^{-3}$, diameter $ 0.99\,\text{m}$ and thickness $d = 0.033 \,\text{m}$. The floe concentration over the middle $5\,$m of tank was $c_f = 0.39$ and the maximum number of disks in the direction of wave propagation is three. The fresh water in the basin has depth $H = 3.1 \, \text{m}$ and density $\rho = 1000\,\text{kg}\,\text{m}^3$. In the regular wave tests, incoming waves of amplitude $A$ and period $\tau$ are generated. 
For the irregular wave tests, incoming waves have spectrum \eqref{eqn:JS_Norm} with prescribed significant height $H_s$ and peak period $\tau_p$. 
Table~\ref{tab:exp} summarises the incoming wave properties over all tests conducted. Throughout the tests, overwash is observed as a consistent build-up of water on the surface of the disks. 
Overwash of the array of disks is summarised by counting the number of overwashed disks along a three-disk transect in the direction of wave propagation along the centre of the basin (black dashed rectangle). 

The disks are modelled using the length $L = 0.99\,\text{m}$ (i.e.\ disk diameter), Young's modulus $E = 4 \, \text{GPa}$ and Poisson ratio $\nu= 0.3$ \cite{Bennetts-2015}. The model is used to predict the maximum extent $X_{d,L}$ of overwash for a three floe transect in the degenerate case where all floe properties and their respective counts are known. The results are reported as the number of floes overwashed $M_X$, which is calculated as $M_X = cX_{d,L}/L$. 
The wave dissipation function $\mathcal{T}_{diss}$ is set to unity, as attenuation due to scattering dominates for the wavelength to floe length ratios in the experiment \cite{Bennetts-2015,Toffoli-22}.

Figures~\ref{Fig:OW_PRSA}(b) and \ref{Fig:OW_PRSA}(c) show the model predictions of maximum extent for incoming waves over ranges of $A$ and $\tau$ for regular waves (Figure~\ref{Fig:OW_PRSA}b) and $H_s$ and $\tau_p$ for irregular waves (Figure~\ref{Fig:OW_PRSA}c). Each combination of the values defines an incoming wave (regular) or an ensemble of waves (irregular), from which the number of floes overwashed ($M_X$) is predicted. The experimental observations are overlaid for each test conducted, 
and the model predictions and experimental observations agree well. The tests provide a range of overwash behaviours from no disks overwashed to all disks overwashed, all of which the model predicts accurately.

The regular wave results in Figure \ref{Fig:OW_PRSA}(b) provide insight into the effect of the incoming wave properties on the extent of overwash. As expected, increasing the wave amplitude increases the extent of overwash into a floe field. Increasing the wave period increases extent at first and then begins to decrease extent. The period dependence is a result of the relative wavelength (period/frequency) dependence of the floe--wave motions discussed in \S\ref{subsec:Floe--Wave}. Larger reflection of energy makes individual floes easier to overwash but decreases the transmission of wave energy farther into the floe field. When wavelengths are small compared to floe thickness and length (small periods) most energy is reflected and the floe is approximately still. The large reflection and negligible movement of the floe results in the first disk overwashing with the least incoming amplitude (energy), which corresponds to $A \approx d_f/2$. Overwashing subsequent disks requires ever larger incoming amplitudes (energy) as most wave energy is reflected and thus not transmitted. In contrast, when wavelengths are long compared to floe thickness and length (long periods) then most energy is transmitted. The significant transmission of waves causes the first floe to require greater incoming amplitudes (energy) to overwash as wave reflection is reduced, but, once achieved, overwash extends deep into the floe field.
The large transmission of wave energy for longer periods becomes extreme for $\tau > 1 \, \text{s}$, resulting in predictions of either no disks overwashing or all disks overwashing. Due to the competition between reflection and transmission, each overwash extent has a period at which the incoming amplitude that achieves the extent is minimised, and this period and its associated amplitude increase as the overwash extent does. 

The irregular wave model predictions of overwash extent (Figure~\ref{Fig:OW_PRSA}c) demonstrate dependencies on $H_s/2$ and $\tau_p$, which are qualitatively similar to the dependencies on $A$ and $\tau$ for the regular wave predictions. The behaviour for short $\tau_p$ is almost identical to short $\tau$ regular waves as most of the wave energy lies in frequencies (wavelengths) in the large reflection regime, meaning overwash requires smaller incoming $H_s/2$ but does not penetrate far into the floe field. 
The similarity between the results for irregular and regular waves extends to the minimum $H_s/2$ 
and corresponding $\tau_p$ (or $A$ and $\tau$) required to produce an overwash extent. As $\tau_p$ (or $\tau$) increases, the overwash extent produced by $H_s/2$ (or $A$) is greater for the irregular waves than the regular waves. 
This is because all the frequency components of the irregular waves contribute to the coupled floe--wave motions, which can drive overwash. Further, the condition on overwash of irregular waves $f_o > f_{tol} = 0.05$ is weaker than in the regular wave case, where either $f_o = 1$ or $f_o = 0$. The presence of multiple wave frequencies for irregular waves also results in the sharp transition between all disks overwashing or no disks overwashing observed in the regular wave results for $\tau > 1 \, \text{s}$ (Figure \ref{Fig:OW_PRSA}b) being replaced by a more gradual transition in $M_X$ for $\tau_p > 1 \, \text{s}$. Both regular and irregular wave overwash extent predictions do not demonstrate effects from resonant responses in the coupled--floe wave motions, as $M_X$ contours are smooth and there are no sudden peaks in $M_X$ for particular $\tau_p$ (or $\tau$). The lack of resonant effects for overwash extent is explained by the lack of these effects on individual floe overwash predictions as demonstrated in Figure~\ref{Fig:OW_Q_Ex}, and also on attenuation due to the floe field as demonstrated in Figure~\ref{Fig:OW_Q_Trans}.
\begin{table}
 \centering
 \begin{tabular}{c| l l l l l l l | l l l}
 $\tau$ or $\tau_p$ $[\text{s}]$ & $0.65$& $0.95$& $0.95$& $1.25$&$1.25$&$1.55$&$1.85$&$0.8$&$1.4$&$1.4$ \\
 \hline
 $A$ or $H_s /2$ $[\text{m}]$ & $0.01$& $0.015$& $0.03$& $0.02$& $0.04$& $0.02$& $0.02$& $0.01$& $0.02$& $0.04$ \\
 \end{tabular}
 \caption{Incoming waves properties for the regular (first group) and irregular (second group) wave experiments.}
 \label{tab:exp}
\end{table}
\section{Predictions of overwash extent in the MIZ}\label{sec:predictions}
The extent of overwash for fragmented \cite{Toyota-2011} and pancake \cite{Alberello-2019} floe fields described in \S\ref{subsubsec:FSD} is investigated. 
The floe fields are assumed to have a compact ice edge after which the FSD and floe concentration are constant. The studied spectra are the JONSWAP spectra \eqref{eqn:JS_Norm} with various values of $H_s$ and $\tau_p$ prescribed. 
The predicted extent of overwash for a section of the Antarctic MIZ is then produced using the floe field and incoming wave properties around the South African icebreaker S.A.~Agulhas~II on a July 2017 voyage, during which wave evolution and the FSD in the MIZ were monitored by an onboard stereo-camera system \cite{Alberello-2019,Alberello-2021}. The MIZ observed by the S.A.~Agulhas~II had a diffuse ice edge, which is accounted for in the predictions using satellite derived ice concentrations. 

\subsection{Overwash Extent for Floe with Given Properties}
\subsubsection{Effect of Incoming Wave Spectra}
\label{subsubsec:EffIncWave}
\begin{figure}
 \centering
 \includegraphics[width=\textwidth]{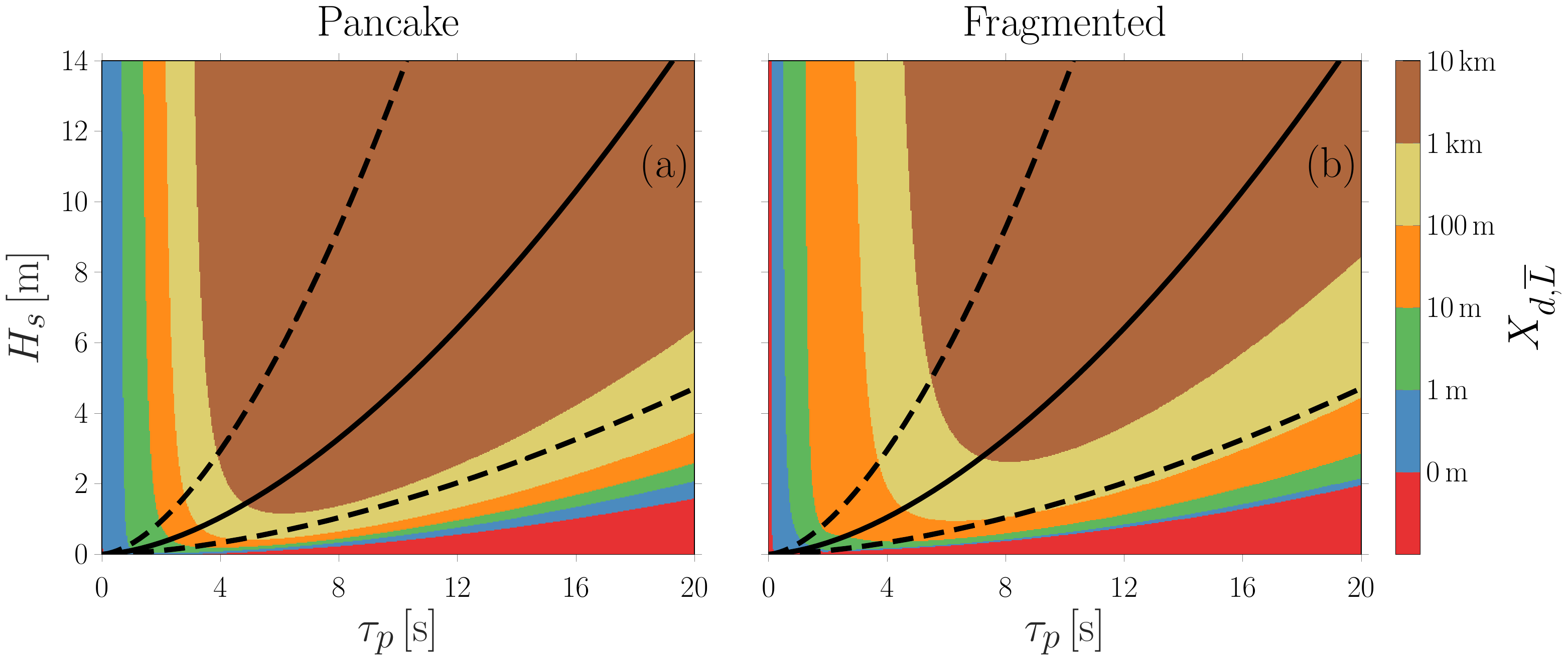}
\caption{Effect of incoming wave spectra on maximum overwash extent for a particular floe ($L = \overline{L} = 0.7\,\text{m}$) in a (a)~pancake floe field \cite{Alberello-2019} and a particular floe ($L = \overline{L} = 5\,\text{m}$) in a (b)~fragmented floe field \cite{Toyota-2011}. Relationship \eqref{eqn:Energy_rel} is overlaid ({\color{black}\solidrule[2mm]}), as well as \eqref{eqn:Energy_rel} scaled ({\color{black}\dashedrule}) to bound wave data in the Southern Ocean \cite{Young-2020}. }
\label{Fig:OW_WP}
\end{figure}

Figure~\ref{Fig:OW_WP} shows the dependence of the maximum overwash extent for floes with thickness $d$ and length $L$, i.e.\ $X_{d,L}$, on the incoming $H_s$ and $\tau_p$, 
for (a)~pancake and (b)~fragmented floe fields. 
The chosen floe lengths are the means for the FSDs, which are $\overline{L} = 0.7 \, \text{m}$ for pancake floes and $\overline{L} = 5 \, \text{m}$ for the fragmented floes. 

Figures~\ref{Fig:OW_WP}(a--b) demonstrate qualitatively similar relationships between overwash extent and the incoming waves ($H_{s}$ and $\tau_{p}$) for the pancake and fragmented floe fields.
They both show that increasing $H_s$ increases overwash extent, and that increasing $\tau_p$ at first increases overwash extent and then decreases overwash extent past some critical peak period.
The relationships match those found in the experimental validation (Figure~\ref{Fig:OW_PRSA}c), indicating that these relationships hold despite the inclusion of wave dissipation and the random variation of floe length according to the FSD. 

In the Southern Ocean, mean values of $H_s$ are between $3\,\text{m}$ and $6\,\text{m}$ \cite{Young-2020}, with measured values up to
$H_s=14\,\text{m}$
\cite{Young-2020}. 
Peak periods and significant wave heights tend to exhibit the following mean relationship \cite{Young-2020}
\begin{equation}
 H_s = 4 \sqrt{6.36531026 \times 10^{-6} u_{10}^{0.7} g^{1.3} \tau_p^{3.3}},
 \label{eqn:Energy_rel}
\end{equation}
where $u_{10}=12 \, \text{m} \, \text{s}^{-1}$ is the annual mean wind speed $10\, \text{m}$ above the ocean surface \cite{Young-2020}. 
The relationship \eqref{eqn:Energy_rel} is overlaid on Figures~\ref{Fig:OW_WP}(a--b), as well as scaled versions that bound the available wave data \cite{Young-2020}. 

Considering only the $H_s$ and $\tau_p$ inside the bounding relationships (dashed lines), provides predictions of typical overwash extents for pancake and fragmented floe fields. For the pancake floe field (Figure~\ref{Fig:OW_WP}a), the maximum overwash extent for the chosen mean floe in mean seas ($H_s = 3\text{--}6 \, \text{m}$) is between $100 \, \text{m} \text{--}10 \, \text{km}$. For mean seas extents are typically over $1 \, \text{km}$ with the exception being those with comparatively long $\tau_p$ given \eqref{eqn:Energy_rel} (bottom dashed line). Maximum extents are also less than $1 \, \text{km}$ for calm seas $H_s \le 2 \, \text{m}$. For the fragmented floe field in mean seas, typically $100\, \text{m}<X_{d,\overline{L}}<10\, \text{km}$. Overwash extents over $1\, \text{km}$ occur for mean seas with corresponding $\tau_p$ approximately given by \eqref{eqn:Energy_rel}, whereas overwash extents below $1\, \text{km}$ occur for seas with $H_s < 3 \, \text{m}$ and those with spectra around the lower bound relationship between $H_s$ and $\tau_p$. Both fragmented and pancake floe fields have $X_{d,\overline{L}}<10 \, \text{km}$ for even the most energetic seas. 

Comparing the predictions of both types of floe field, overwash penetrates less into fragmented floe fields, requiring about twice the $H_s$ values to generate $X_{d,\overline{L}} \ge 1\, \text{km}$. The reduction of overwash extents in fragmented floe fields compared to pancake floe fields is due to the floes being longer and thicker. Fields of thicker and longer floes reflect more wave energy decreasing the extent of overwash. Thicker floes also have larger freeboards making them more difficult to overwash.

\subsubsection{Effect of Floe Properties}
\label{subsubsec:EffFloeProp}

\begin{figure}
 \centering
 \includegraphics[width=\textwidth]{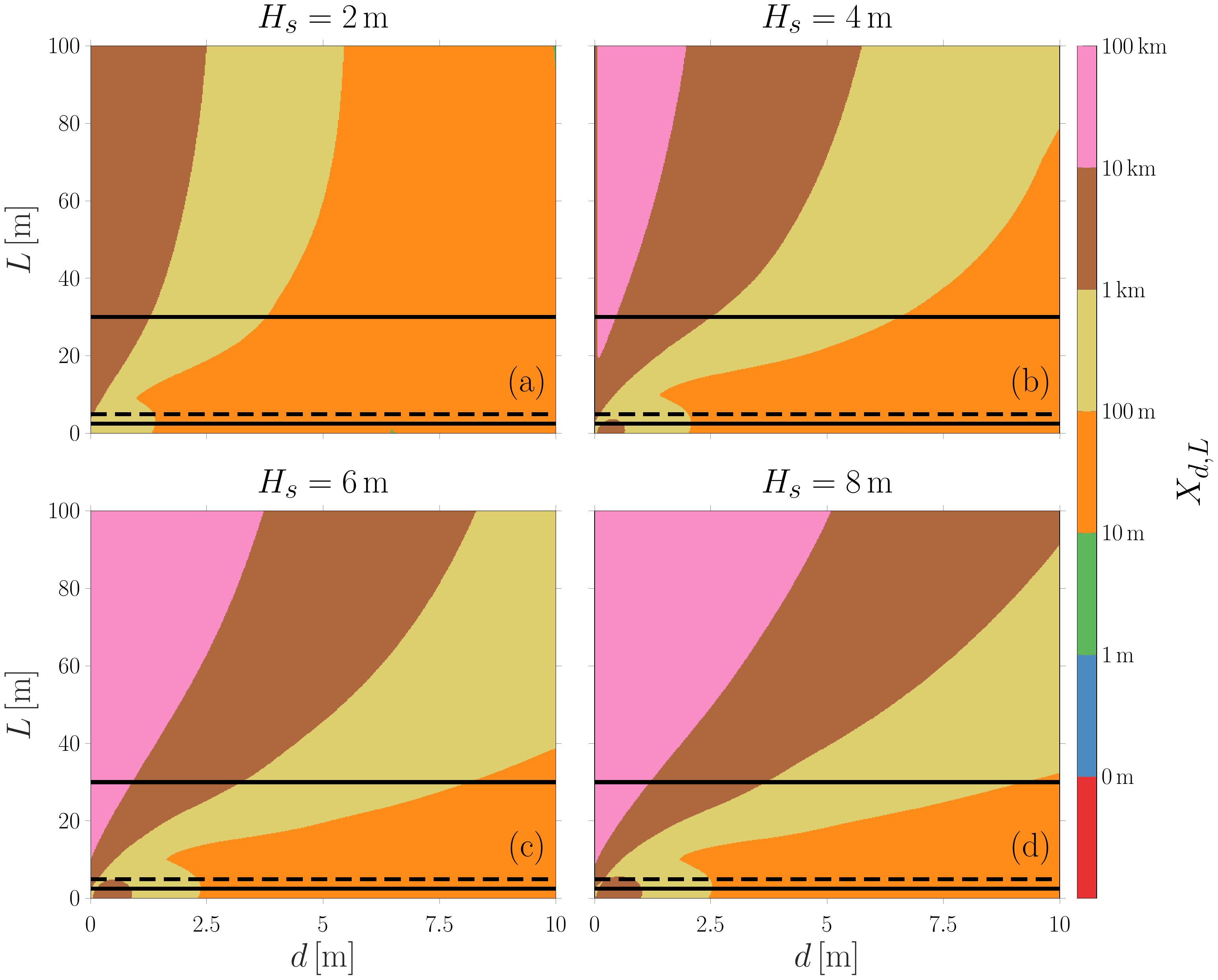}
\caption{Effect of particular floes length $L$ and thickness $d$ on maximum overwash extent for particular floe sitting in a fragmented floe field \cite{Toyota-2011} with a concentration of $c_f= 0.6$. The FSD parameters of $L_{crit}$ and $L_{min}$ (\solidrule{}) as well as $\overline{L}$ (\dashedrule{} ) are demonstrated. Incoming ocean wave spectra with a range of $H_s$ values (a)-(d) and $\tau_p$ determined by \eqref{eqn:Energy_rel}.}
\label{Fig:OW_IP}
\end{figure}

Figure~\ref{Fig:OW_IP} shows the dependence of the maximum overwash extent for a floe in a fragmented floe field on the length and thickness of that floe. 
Results are shown for four different values of $H_{s}$, with corresponding values of $\tau_{p}$ such that relationship \eqref{eqn:Energy_rel} is satisfied. 
Changes in floe length affect the chosen floe only, with the lengths of all other floes in the floe field determined by the FSD. 
The mean floe size, $\overline{L}$, is indicated on the plots, along with $L_{min}$ and $L_{crit}$, with $97\%$ of floes having lengths between these values.
In contrast, the thickness of the floe affects the entire floe field, as a uniform thickness for the floe field is assumed. 
Thickness values of order $1$\,m are expected in a fragmented floe field \cite{Toyota-2011}, and so a thickness of $10\,\text{m}$ provides an upper bound.

Figures~\ref{Fig:OW_IP}(a--d) show similar qualitative relationships between maximum overwash extent and floe length and thickness.
Increasing floe thickness decreases overwash extent, as increasing thickness increases freeboard and wave attenuation, making overwash of individual floes more difficult and reducing overwash extent. Increasing floe lengths increases maximum overwash extent. Longer floes can be overwashed by longer waves (lower frequency) due to increased wave reflection. Longer waves (lower frequency) travel further into the floe field since the long floes that scatter them are rare in the split power law FSD and so longer floes have larger overwash extents. When the floe lengths and thicknesses are small and of comparable size, there are regions of $100 \, \text{m}$ and $1 \, \text{km}$ overwash extent that expand with increasing $H_s$.
These growing regions are a result of a small but consistent amount of overwash of thin floes (with small freeboards) due to the presence of multiple low frequency (large wavelength) components. Individually, such low frequency components would not overwash the floe, but together they do, and since these are low frequency (large wavelength) waves, they propagate far into the floe field. 
The contour is growing as incoming $H_s$ increases as the energy in these frequency components grows. 


In calm seas ($H_{s}=2$\,m; Figure~\ref{Fig:OW_IP}a), most floes ($L_{min} < L< L_{crit}$) with $d \le 1\, \text{m}$ will be overwashed $0.1 \, \text{--} 1 \, \text{km}$ into the MIZ, with almost all floes in floe fields with $d \ge 5\, \text{m}$ only being overwashed in at least the first $10 \, \text{m}$. For mean to high seas (Figures~\ref{Fig:OW_IP}b--d) in floe fields with $d \le 1\, \text{m}$ extents of $1 \, \text{km}$ occur for relatively long floes ($\overline{L} < L< L_{crit}$). Extents of $10 \, \text{km}$ are also possible for higher seas (Figures~\ref{Fig:OW_IP}c--d) and long floes ($L > 20 \, \text{m}$) in thin floe fields ($d < 0.5 \, \text{m}$). Extents for floe fields with $d \ge 5\, \text{m}$ typically remain less than $100 \, \text{m}$, with the exception of relatively long floes ($\overline{L} < L< L_{crit}$) for which extents over $100 \, \text{m}$ are possible. The maximum overwash extent for relatively and extremely long floes ($ L > \overline{L}$) is most sensitive to increasing $H_s$, as increasing $H_s$ also increases $\tau_p$, meaning more wave energy at longer periods (wavelengths), which drives the overwash of these long and rare floes. For short ($L< \overline{L}$) and thick ($d> 2.5 \, \text{m}$) floes, increasing $H_s$ does not noticeably increase overwash extent. 

\subsection{Expected Overwash Extent for Floe Field}
\label{subsec:ProbExt}
\begin{figure}
 \centering
\includegraphics[width=\textwidth]{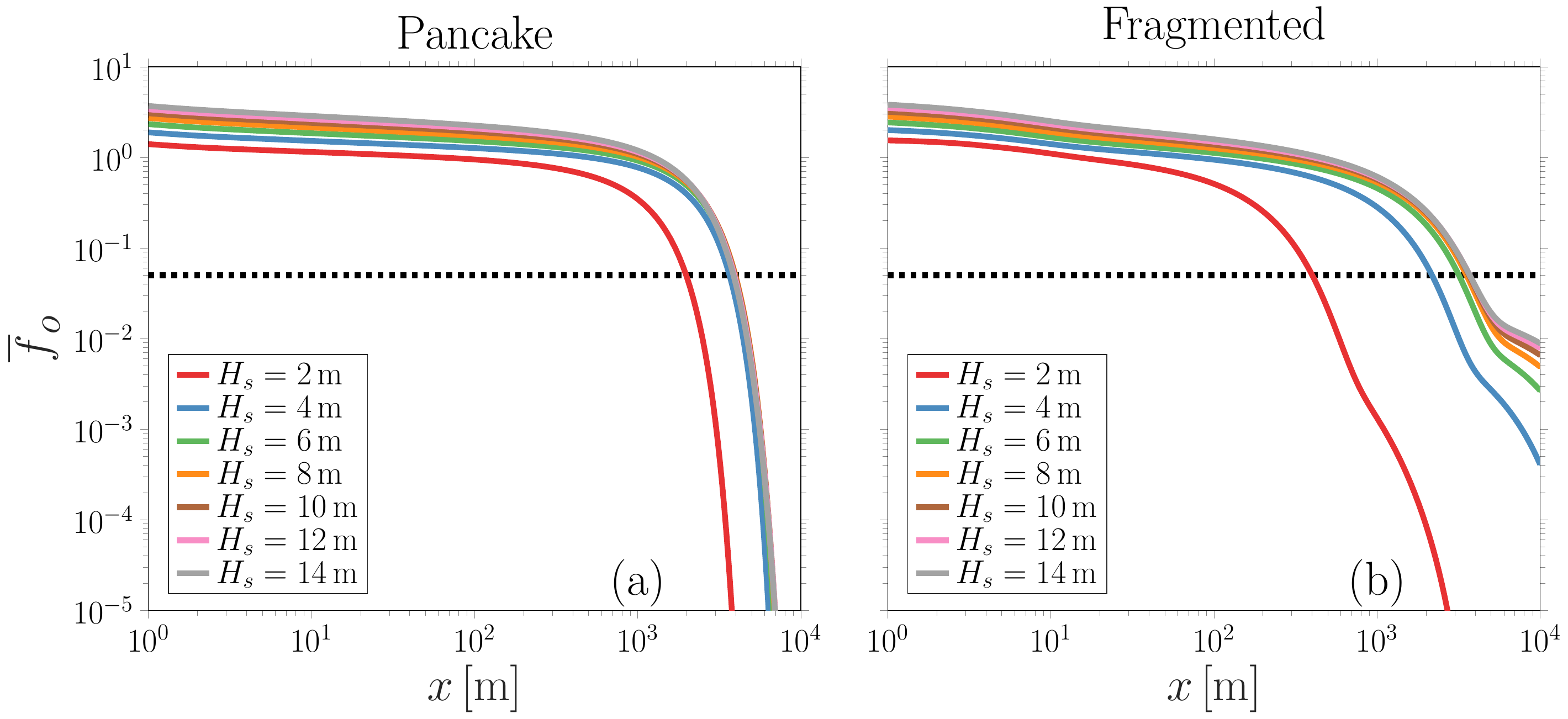}
 \caption{Expected overwash event frequency over distance ($\overline{f}_o(x)$) for (a)~pancake and (b)~fragmented floe fields and various $H_s$ values where $\tau_p$ is given by \eqref{eqn:Energy_rel}. The tolerance frequency $f_{tol}$ (\dashedrule{}) is used to determine the expected extent $\overline{X}_d$.}
 \label{Fig:OW_OverwashFrequency}
\end{figure}

Figure \ref{Fig:OW_OverwashFrequency} shows the expected overwash event frequency ($\overline{f}_{o}$) versus distance, for (a)~a pancake, and (b)~a fragmented floe field, for
a range of incoming wave spectra \eqref{eqn:JS_Norm} with various $H_s$-values where $\tau_p$ is given by \eqref{eqn:Energy_rel}. 
The value of $f_{tol} =0.05$ is highlighted to demonstrate the expected overwash extent ($\overline{X}_d$) for each incoming spectrum. 

Figures~\ref{Fig:OW_OverwashFrequency}(a--b) demonstrate similar behaviour of $\overline{f}_{o}$ over distance. In particular, $\overline{f}_{o}$ decreases with increasing distance due to attenuation of the waves driving overwash. The apparent decrease is gradual at first and then takes place rapidly after a kilometre. The gradual and then rapid reduction in $\overline{f}_{o}$ is caused by the exponential attenuation of the wave field and enhanced by the logarithmic scale of the plots. Increasing $H_s$ increase $\overline{f}_{o}$, as expected.

The pancake floe field in calm seas ($H_s = 2 \, \text{m}$; red line) is predicted to have most floes experiencing an overwash event about once
every mean period ($\overline{f}_{o}\approx{}1$) 
up to a distance of $x=100\, \text{m}$. 
As distance increases, $\overline{f}_o$ decreases due to attenuation removing wave energy from the frequencies that drive overwash of most floes. 
After a kilometre, the decrease becomes rapid, and the overwash frequency drops by several orders of magnitude. 
The overwash frequency drops below $f_{tol} =0.05$ when $x=\overline{X}_{d}=1.9$\,km resulting in an attenuated $H_s = 0.79 \,\text{m}$. The overwash frequency $\overline{f}_o$ (and thus $\overline{X}_d$) increases with increasing $H_{s}$, but maintains the same qualitative behaviour. The increase is greatest when $H_s$ increases from $ 2\, \text{m}$ to $4\, \text{m}$ ($\overline{X}_{d}=3.6$\,km where $H_s = 2.3 \,\text{m}$ ), but has less effect as $H_s$ increases further, for example by doubling again to $H_s =8\, \text{m}$ ($\overline{X}_{d}=3.9$\,km where $H_s = 6.2 \,\text{m}$ ). The reduced effect of doubling $H_s$ is due to the exponential nature of attenuation and the associated increase in $\tau_p$ \eqref{eqn:Energy_rel}. The increase in $\tau_p$ removes energy from the short period (high frequency) waves that drive overwash, as demonstrated by the prediction of no overwash when attenuated $H_s$ values are large. Therefore, in calm seas most floes in the pancake floe field will overwash at a distance of $1.9 \,\text{km}$, with higher seas ($H_s > 4\,$\text{m}) allowing overwash up to distances over $3\, \text{km}$.

The average overwash extent of fragmented floe fields is more sensitive to $H_s$ with calm seas ($H_s = 2 \, \text{m}$) resulting in $\overline{X}_{d}=400$\,m (attenuated $H_s = 0.8 \,\text{m}$), and extremely high seas ($H_s = 14 \, \text{m}$) resulting in $\overline{X}_{d}=3.7$\,km (attenuated $H_s = 12 \,\text{m}$). This greater sensitivity of the fragmented floe field is due to the greater variability in the floe lengths in its FSD. Of particular importance is the larger presence of longer floes in the fragmented floe field for which overwash is more likely. The shape of $\overline{f}_o$ over distance is different for fragmented floe fields (although the overall trends are the same), in particular, the second region of gradual decrease in $\overline{f}_o$ around the value of $f_{tol}$. This shape difference is due to the increased role of scattering on attenuation for the fragmented floe field as more floes have lengths and thickness comparable to the wavelengths investigated.

\subsection{Overwash Extents vs.\ Field Observation Locations}
\label{subsec:Agulhas}

On the $4^{\text{th}}$ of July $2017$, the South African icebreaker S.A.~Agulhas~II entered the Antarctic MIZ with a polar cyclone nearby \cite{Vichi-2019,Alberello-2019,Alberello-2021}. An onboard stereo-camera system was used to monitor waves and floe sizes in the MIZ over 30\,minute intervals at six locations as the S.A.~Agulhas~II progressed deeper into the MIZ up to 40\,km from the ice edge.
Analysis of the floe sizes by \citeA{Alberello-2019} produced the pancake floe field FSD described in \S\ref{subsubsec:FSD}.
Maximum wave heights greater than $9\, \text{m}$ were recorded over $20 \, \text{km}$ into the MIZ \cite{Alberello-2021}. 

\begin{figure}
\centering
\includegraphics[width=\textwidth]{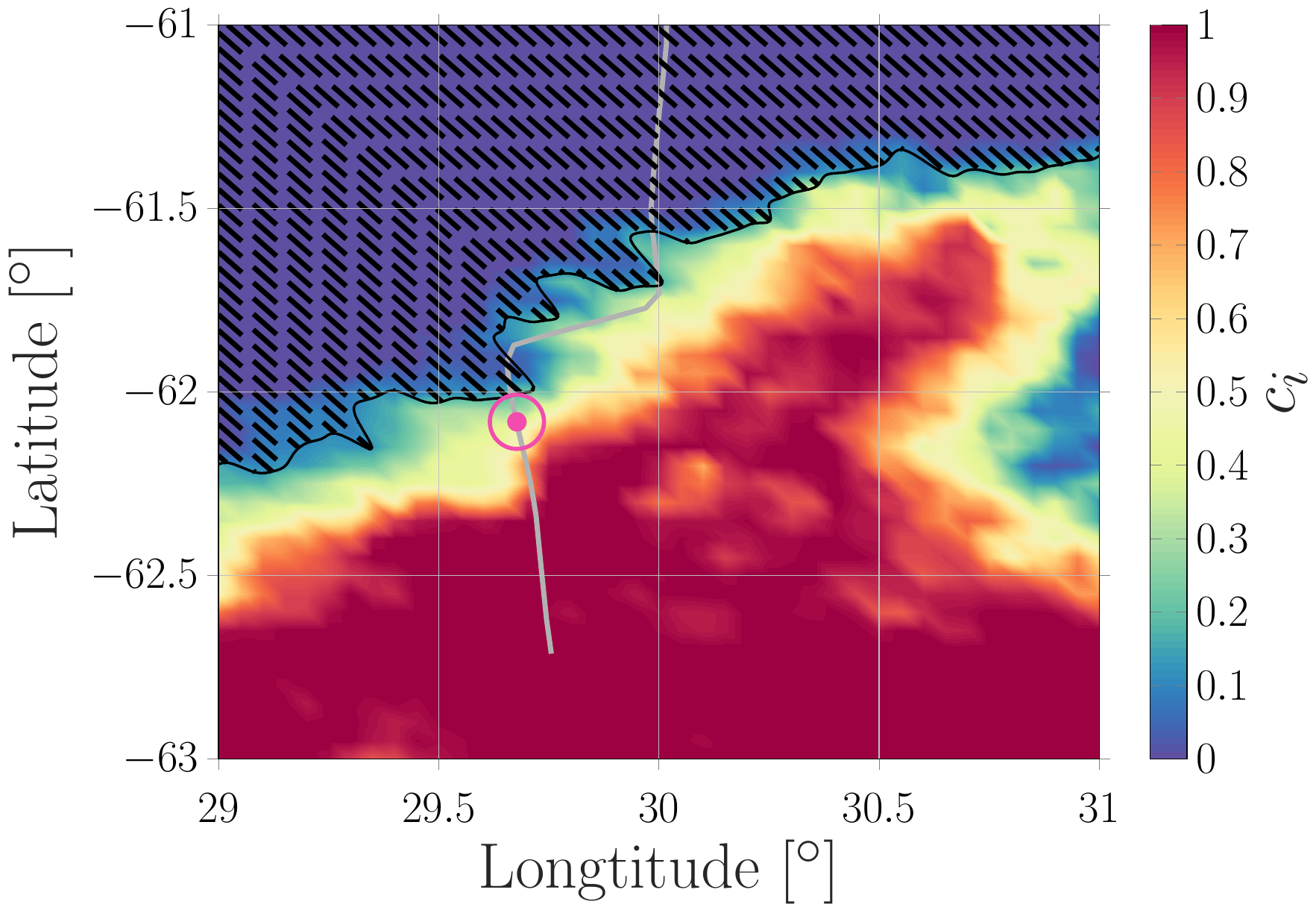}
\caption{ Plot of the ice concentration $c_i$ from AMSR2 satellite data \cite{Beitsch-2014}, overwashed portion of transects ({\color{black}\dashedrule}), predicted overwash region ({\color{black}\solidrule[2mm]}) and the path of the S.A.~Agulhas~II ({\color{grey_1}\solidrule[2mm]}) and the reported location of first measurement (\ColorCircle{magenta!70!white}). The distance travelled during the first wave measurements (\BoundCircle{solid}{1pt}{magenta!70!white}{white}) which took place between 08:00 and 08:30 UTC on the $4^{th}$ of July $2017$.}
\label{Fig:Arg_Trans}
\end{figure}

Figure \ref{Fig:Arg_Trans} shows the average ice concentration $c_i$ during the day of the voyage from the AMSR2 satellite data \cite{Beitsch-2014} and the incoming track of the S.A.~Agulhas~II, which begins at $-61^\circ$ latitude and continues up to $-62.7^\circ$. 
The first series of stereo-camera images were acquired between 08:00 and 08:30\,UTC, and the locations along the track during this period are indicated in terms of the location at 08:15\,UTC (pink bullet) and the entire section of the track over the 30\,minutes (intersection of the circle with the track).
Model predictions of expected overwash extent ($\overline{X}_d$) along transects in the mean wave direction are overlaid, and these transects are combined to produce a prediction of the region in which most floes in the floe field will be overwashed. The incoming wave spectra for the transects are the JONSWAP \eqref{eqn:JS_Norm} where $H_s$ and $\tau_p$ values are given by the ECMWF ERA5 wave reanalysis \cite{Hersbach-2020} at 08:00 UTC at the $61^{\text{st}}$ parallel south. The ice concentration $c_i$ derived from AMSR2 satellite data \cite{Beitsch-2014} which combines floe and interstitial ice concentrations is used as an estimate for the floe concentration $c_f$. While $c_f$ was produced along the ship track well into the MIZ \cite{Alberello-2019}, the limited spatial extent of this concentration in particular the lack of ice edge information makes this $c_f$ insufficient for the entire investigated region. 

The predictions of expected overwash extent demonstrate that the first series of stereo-camera images commenced very close to the location at which overwash becomes unlikely to occur, and, hence all subsequent images are beyond the overwash region. Overwash remains restricted to the outermost region of the MIZ (taken to be where $c_i > 0.01$), with a farthest penetration of $16\,\text{km}$ where ice concentration remained low over significant distances between $29^\circ$ and $29.5^\circ$ east longitude. The highest ice concentration where overwash is predicted to occur is $c_i \approx 0.4$, which occurred only $7\,\text{km}$ into the MIZ at around $61.5^\circ$ south latitude. Inside the predicted region of overwash, the model indicates that most floes will overwash at least once every $20$ mean periods ($\overline{f}_o > 0.05$). In the overwash region $\overline{f}_o$ has mean value $1.3$ and so most floes have an overwash event about $1.3$ times every mean period. The mean is much higher than $f_{tol}$, as that is the cut-off and because $\overline{f}_o$ reduces quite rapidly over long distances as shown in Figure~\ref{Fig:OW_OverwashFrequency}. Using the mean periods measured during the voyage, $\overline{\tau}_I(0) \approx 12 \, \text{s}$ \cite{Alberello-2021}, an observer on a ship should expect to see overwash events about once every $16$\,s and at least once every $2.5$\,minutes.

\section{Conclusions and Discussion}\label{sec:conclusions}
Wave overwash extents of ice floes in the MIZ have been generated using a novel model that incorporates the stochastic nature of the irregular incoming waves and floe sizes. The model produces the expected frequency of overwash events indicating the frequency of overwash events for most floes in the MIZ. Model predictions were given for pancake floe fields, based on measurements from the Antarctic MIZ during the winter ice advance \cite{Alberello-2021}, and fragmented floe fields generated by breakup of a continuous ice cover,
based on measurements from the Weddell Sea at the beginning of the spring ice retreat \cite{Toyota-2011}. 
For mean sea states, with significant wave heights $2$--$6$\,m and associated peak periods, most floes were overwashed $1.9\text{--}3.9\, \text{km}$ into the pancake floe field, at which distances $H_s$ had attenuated to $0.8$--$4.2$\,m. For a fragmented floe field in mean seas, most floes were overwashed up to $0.4\text{--}3.1\, \text{km}$, by which distances $H_s$ had attenuated to $0.8$--$4.2$\,m. 
The limited extent of predicted overwash into both floe fields and relatively large waves needed to cause overwash helps to explain the scarcity of in situ reports of overwash in the MIZ.

Overwash extent was larger for the pancake floe field than the fragmented floe field, as pancake floes were smaller and thinner. All else being equal (i.e.\ incoming waves and ice concentrations), this indicates that overwash extents are greater during during autumn--winter (when ice is advancing and the MIZ is dominated by pancake floes) than spring--summer (when ice is retreating and the MIZ is dominated by fragmented floes). Further, waves in the Southern Ocean are typically most energetic in winter \cite{Young-2020}, which will produce greater overwash extents. Seasonal changes in ice conditions such as floe sizes \cite{Nihashi-2001} and ice concentrations \cite{Cavalieri-2008} tending to increase in autumn--winter reducing overwash extent and decreasing in spring--summer increasing overwash extent are also a factor. However, these general trends in ice conditions may be less apparent at the outermost MIZ where overwash is predicted to occur as the outermost MIZ moves in response to these seasonal changes. 

The overwash extent for incoming waves and a floe field representative of the conditions during an experiment in the winter Antarctic MIZ that coincided with a polar cyclone were produced. The experiment used a stereo-camera system \cite{Alberello-2019,Alberello-2021}, which could have, in principle, captured overwash events but did not.
The model was shown to predict that in these condition overwash is limited to the outskirts of the MIZ where ice concentrations are $<0.4$, and before the first series of images were acquired.
Therefore, the predictions suggest that prospective field studies targeting overwash, using a stereo-camera system or equivalent, will need to take place closer to the ice edge. 
In this case, the model predictions from \S\ref{subsec:ProbExt} indicate that measurements could be taken during mean sea state conditions (i.e.\ not during a cyclone), 
as overwash would still persist over several kilometres from the ice edge.

The predicted overwash extents, do not incorporate wave dissipation due to overwash and thus should be interpreted as upper bounds. While models of overwash dissipation for incoming regular waves have been developed \cite{Nelli-2020,Skene-2021}, the extension of these models to irregular waves is an open question. Despite the omission of overwash dissipation the agreement between predictions and laboratory experiments was good. The predictions of overwash extent are based on average overwash behaviour for the ensemble of possible floe fields and incoming waves. Therefore, specific instances of floe fields and incoming waves significantly different to the average of their respective ensembles (according to the underlying probability distribution) may produce dissimilar overwash behaviour. For instance, while predictions of overwash are given in terms of $H_s$ very large individual waves, with heights up to $26.4\,$m (in a record with $H_s=15.6$\,m) have been recorded in the open Southern Ocean \cite{Young-2020} which may generate overwash far deeper into the MIZ.

The model assumes that wave (incoming spectrum) and ice conditions (FSD and ice concentration) are steady. Since the criterion defining overwash ensures that at least one overwash event occurs every twenty mean periods, this is the natural time scale over which the spectrum and FSD should not change for the model to be valid. Since mean periods in the MIZ are of the order of tens of seconds \cite{Alberello-2021,Meylan-2014}, the assumption that waves and ice conditions do not change over the order of minutes to tens of minutes is reasonable. However, the model is not valid for transient waves, such as those generated by a passing ship, for example the observations of \citeA{Dumas-2021PP}.
Additionally, it is assumed that the FSD is constant over distance in the MIZ although ice concentrations are permitted to change over distance. Therefore, banding and diffuse edges of the MIZ can be incorporated insofar as they only vary ice concentration. With the lower ice concentrations associated with banding and diffuse ice edges resulting in increased overwash extents compared to a compact ice edge. 

In conclusion, a model has been developed that predicts the extent of wave overwash of ice floes in the MIZ for specified incoming waves and floe fields.
Model outputs, for the degenerate case where the floe thicknesses and lengths are known, were validated using the only available data from a laboratory wave basin experiment involving an array of artificial floes. Overwash is a critical component to the evolution of the MIZ as it dissipates wave energy, can remove snow from the surface of floes \cite{Massom-1998} and deposits water on a floes surface. The resultant water on a floes surface being likely to enhance both growth \cite{Doble-2003} and melt \cite{Massom-2001} of floes and hosts biota \cite{Ackley-1994}. Overwash and its resultant attenuation and thermodynamic and biogeochemical effects are currently absent from models of sea ice. The proposed overwash extent model can be integrated within current sea ice models \cite{Hunke-2010}, provides a basis for an overwash dissipation model for floe fields and can be used to guide future field observations of overwash.

\section*{Open Research}
The data sets used for \S\ref{subsec:Agulhas} are available online with ice concentration provided by ASMR2 satellite data \cite{Beitsch-2014} and wave conditions provided by ERA5 \cite{Hersbach-2020}.

\acknowledgments
This work is funded by the Australian Research Council (DP200102828).
LGB is supported by an Australian Research Council mid-career fellowship (FT190100404).
RM is supported by the Australian Antarctic Division and the Australian Government’s Australian Antarctic Partnership Programme, and this study contributes to AAS Projects 4116 and 4528 and the Australian Centre for Excellence in Antarctic Science.
The authors thank Alberto Alberello for providing data used in Figure~\ref{Fig:Arg_Trans}. The authors also thank Francois P\'{e}tri\'{e}, Vincent Lafon, Thierry Rippol and Alexandre Cinello (Oc\'{e}anide, La Seyne Sur Mer) for helping design and conduct the experimental campaign analysed in \S\ref{sec:validation}.

\appendix

\section{Overwash Simulation}
\label{App:Sec_SWE}
Given an incoming spectrum $S$, a realisation of the incoming waves from \eqref{eqn:IrregularWave_Components} that is a sample of the random phases $\theta_n$ (uniform-distributed) and the random amplitudes $A_n$ (Rayleigh-distributed with mean $\sqrt{2 S(\omega_n)\Delta \omega}$) is produced.
For each regular wave component of this incoming realisation, the propagating solutions of \citeA{Bennetts-2007} produce the reflection ($R(\omega)$) and transmission ($T(\omega)$) coefficients as well as the plate movement coefficients for each location $x$ ($\hat{\zeta}(x,\omega)$). From these coefficients, the left and right wavefields as well as the plate movements are given by
\begin{align*}
 \eta_l(x,t) &= \sum_{n=1}^{N} A_{n} \Re\left\lbrace (1 + R(\omega_n) e^{-2i k_n x }) e^{i(k_n x - \omega_n t + \theta_n)}\right\rbrace,\\
 \eta_r(x,t) &= \sum_{n=1}^{N} A_{n} \Re\left\lbrace T(\omega_n) e^{i(k_n x - \omega_n t + \theta_n)} \right\rbrace, \\
 \zeta(x,t) &= \sum_{n=1}^{N} A_{n} \Re\left\lbrace \hat{\zeta}(x,\omega_n) e^{i(k_n x - \omega_n t + \theta_n)} \right\rbrace.
\end{align*}
The velocity at the free surface can also be obtained from these and are given by
\begin{align*}
 u_l(x,0,t) &= \sum_{n=1}^{N} A_{n} \frac{g k_n}{\omega_n} \Re\left\lbrace (1 + R(\omega_n) e^{-2i k_n x }) e^{i(k_n x - \omega_n t + \theta_n)}\right\rbrace\quad \text{and}\\
 u_r(x,0,t) &= \sum_{n=1}^{N} A_{n} \frac{g k_n}{\omega_n} \Re\left\lbrace T(\omega_n) e^{i(k_n x - \omega_n t + \theta_n)} \right\rbrace.
\end{align*}

Realisations of the sea surface coupled with the nonlinear shallow water equations (SWEs) for a horizontal bed are used to simulate overwash on the floe surface, as performed by \citeA{Skene-2015}. The nonlinear SWEs for a horizontal bed are 
	\begin{align*}
	&\frac{\partial h}{\partial t} + \dfrac{\partial (uh)}{\partial x} = 0, \\ \nonumber \\
	&\dfrac{\partial (uh)}{\partial t} + \dfrac{\partial}{\partial x} \left ( u^2h + \dfrac{gh^2}{2} \right ) = 0,
	\end{align*}
for $x\in[0,d]$, where $h(x,t)$ is the overwash depth, $u(x,t)$ is its depth-averaged horizontal velocity. 
Homogeneous initial conditions are applied, i.e.\ $h(x,0)=0$ and $u(x,0) = 0$, and overwash is forced by the boundary conditions 
\begin{equation*}
 h(0,t) =\eta_l(0,t),\quad h(d,t) =\eta_r(d,t) ,\quad u(0,t) = u_l(x,0,t) \quad\text{and}\quad u(d,t) = u_r(x,d,t).
\end{equation*}
The initial--boundary-value problem is solved using the numerical method described by \citeA{Skene-2015}, producing the simulation of overwash.

\bibliography{Overwash.bib}

\end{document}